\documentclass[prd, preprintnumbers, showkeys,  nofootinbib, 11pt]{article}
\usepackage[english]{babel}
\usepackage{graphics}
\usepackage{float}
\usepackage[pdftex]{graphicx}
\usepackage[font=small,format=plain,labelfont=bf,up,textfont=it,up]{caption}
\usepackage{bm,bbm}
\usepackage{array}
\usepackage{color}
\usepackage{setspace}
\usepackage{fancyhdr}
\usepackage{amssymb, amsopn, amsmath} 
\usepackage{varioref}
\usepackage{mathrsfs}
\usepackage{lmodern}
\usepackage{url}
\usepackage{xcolor}
\definecolor{dark}{rgb}{0.10,0.2,0.3}
\definecolor{light}{rgb}{1.7,1.5,0.6}
\definecolor{purpure}{rgb}{0.5,0.15,0.3}
\usepackage{hyperref}
\hypersetup{colorlinks,%
citecolor=dark,%
filecolor=dark,%
linkcolor=blue,%
urlcolor=purpure,%
pdftex}

\usepackage{graphicx}

\begin{document}

\newpage
\pagestyle{plain}

\title{ \bf  Gluon Regge trajectory at two loops from Lipatov's high energy effective action
 }

\author{ {  G.~Chachamis$^1$, M.~Hentschinski$^{2}$, J.~D.~Madrigal Mart\'inez$^3$,  A.~Sabio~Vera$^3$}  
\bigskip \\ \\
}
\date{}
\maketitle

\begin{verse}
\vspace{-.5cm}
  { $^1$~Instituto de F\'isica Corpuscular UVEG/CSIC,
E-46980 Paterna (Valencia), Spain}.\\ \vspace{.1cm}
{ $^2$~Physics Department, Brookhaven National Laboratory, Upton, NY 11973,
USA.}\\  \vspace{.1cm}
{ $^3$~Instituto de F\'isica Te\'orica UAM/CSIC, Nicol\'as Cabrera 15 \& }\\ \vspace{.1cm}
{\hspace{.2cm} Universidad Aut\'onoma de Madrid, C.U. Cantoblanco, E-28049 Madrid, Spain.}
\vspace{.5cm}
\end{verse}
\begin{abstract}
  We present the derivation of the two-loop gluon Regge
  trajectory using Lipatov's high energy effective action and a
  direct evaluation of Feynman diagrams. Using a gauge invariant
  regularization of high energy divergences by deforming the
  light-cone vectors of the effective action, we determine the
  two-loop self-energy of the reggeized gluon, after computing the master integrals involved using
the Mellin-Barnes representations technique. The self-energy is further matched to QCD
  through a recently proposed subtraction prescription.  The Regge
  trajectory of the gluon is then defined through renormalization of
  the reggeized gluon propagator with respect to high energy divergences.
  Our result is in agreement with previous computations in the
  literature, providing a non-trivial test of the
  effective action and the proposed subtraction and renormalization
  framework.
\end{abstract}

\section{Introduction}\label{1}

Current applications of high energy factorization to QCD phenomenology
range from the analysis of perturbative observables, such as dijets widely
separated in rapidity \cite{jets}, over transverse momentum dependent
parton distribution functions in the low $x$ region \cite{upd}, up to
the study of phenomena in heavy ion collisions \cite{hic}. Their common base is  the factorization of
QCD scattering amplitudes in the limit of asymptotically large center of
mass energy, together with the resummation of large 
logarithmic contributions using the  Balitsky-Fadin-Kuraev-Lipatov (BFKL)
equation \cite{BFKL1,BFKLNLO}. Recent phenomenological use of the BFKL resummation can be found in the
analysis of the combined HERA data on the structure function $F_2$ and
$F_L$ \cite{Ellis:2008yp, Hentschinski:2012kr}, the study of
di-hadron spectra in high multiplicity distributions at the Large
Hadron Collider \cite{Dusling:2012cg} or the production of high $p_T$ dijets 
\cite{Colferai:2010wu,Ducloue:2013hia,Caporale:2012ih} , widely
separated in rapidity.\\

In the present work we discuss  Lipatov's high
energy effective action \cite{LevSeff} and show that it can serve as a useful tool 
to reformulate the high energy limit of QCD as an effective field
theory of reggeized gluons.  While the determination of the high
energy limit of tree-level amplitudes has been well understood for quite some 
time within this framework \cite{Antonov:2004hh}, it was only until
recently that progress in the calculation of loop corrections has been
achieved.  Starting with \cite{quarkjet} and extended in
\cite{gluonjet}, a scheme has been developed that comprises the 
regularization, subtraction and renormalization of high
energy divergences. This scheme then allowed to successfully derive
forward jet vertices for both quark and gluon initiated jets at NLO accuracy
from Lipatov's high energy effective action. \\ 

Here we extend this program to the calculation of the
2-loop gluon Regge trajectory. The latter provides an essential
ingredient in the formulation of high energy factorization and
reggeization of QCD amplitudes at NLO.  It has been originally
derived in \cite{Fadin:1996tb, Fadin:1995km} using $s$-channel
unitarity relations. The result was then subsequently confirmed in
\cite{Blumlein:1998ib}, clarifying an ambiguity in the  non-infrared divergent contributions of \cite{Korchemskaya:1996je}. The
original result was further verified by explicitly evaluating the high
energy limit of 2-loop partonic scattering amplitudes
\cite{DelDuca:2001gu}. While the explicit result for the 2-loop
gluon Regge trajectory is by now firmly established, our calculation
provides an important confirmation of its universality: unlike
previous calculations, the effective action defines the Regge
trajectory of the gluon without making any reference to a particular
QCD scattering process.\\

For the development of a consistent formulation of the effective action,
the calculation of the 2-loop gluon trajectory provides an essential and non-trivial test of our
scheme. The latter has been set up in \cite{Chachamis:2012gh}, where
partial results, addressing the flavor dependent parts of the gluon
Regge trajectory have been already presented. The current paper addresses the gluon
corrections, which are considerable more complicated than their
fermionic counterparts.\\

The outline of this paper is as follows: Sec.~\ref{2} provides a short
introduction to Lipatov's effective action and a list of necessary
Feynman rules, together with a discussion of our regularization and
the employed pole
prescription. Sec.~\ref{sec:gluontrajectory-effectiveaction} recalls
the scheme we follow in the derivation of the gluon Regge trajectory,
which has been originally introduced in
\cite{Chachamis:2012gh}. Sec.~\ref{sec:computation-two-loop} provides
details about our calculation of the 2-loop reggeized gluon self-energy from the effective action, together with our result for the
2-loop gluon Regge trajectory. Sec.~\ref{5} contains our conclusions
and an outlook on future projects. Several technical details of our
calculations are summarized in the appendix.

\section{Lipatov's  high energy effective action}
\label{2}

The effective action \cite{LevSeff} describes interactions which are
local in rapidity, {\it i.e.} which are restricted to an interval of
narrow width ($\eta$) in rapidity space. The entire dynamics which
extends over rapidity separations larger than $\eta$, is on the other
hand integrated out and taken into account through universal eikonal
factors. To reconstruct from this setup QCD amplitudes in the limit of
large center of mass energies, a new degree of freedom ---the
reggeized gluon--- is introduced on top of the usual QCD fields. The high energy effective action then describes the interaction of this new field with the QCD field content through adding an induced term  $ S_{\text{ind.}}$   to the QCD action $S_{\text{QCD}}$,
\begin{align}
  \label{eq:effac}
S_{\text{eff}}& = S_{\text{QCD}} +
S_{\text{ind.}}, 
\end{align}
where the induced term $ S_{\text{ind.}}$ describes the coupling of the
gluon field $v_\mu = -it^a v_\mu^a(x)$ to the reggeized gluon field
$A_\pm(x) = - i t^a A_\pm^a (x)$. Due to this particular construction,  it is
immediately clear that a specific calculational scheme is needed to
avoid overcounting and to ensure the abovementioned locality in rapidity.
These requirements can be achieved using the following two-step
procedure: a) calculation of vertices of reggeized gluon fields and
QCD degrees of freedom and b) a procedure which matches the resulting
field theory of reggeized gluons with QCD.  a) is achieved through
Lipatov's high energy effective action in Eq.~(\ref{eq:effac}), which provides the gauge
invariant couplings of the new reggeized gluon field to the gluon
field. For b), a certain subtraction scheme has been proposed in
\cite{quarkjet}, originally in the context of quark-quark scattering
at 1-loop, and later on also verified for the
case of gluon-gluon scattering \cite{gluonjet}.\\

 To set the notation it is useful to
have a partonic scattering process $p_a + p_b \to p_1 + p_2 +  \ldots$ in mind with
light-like momenta $p_a^2 = p_b^2 = 0$ and squared center of mass
energy $s = 2 p_a \cdot p_b$.  Dimensionless light-like four vectors
$n^\pm$ normalized to $n^+ \cdot n^- = 2$ are then defined through a
re-scaling $n^\pm = 2 p_{a,b}/\sqrt{s}$, while a general four-vector
$k$ has the decomposition
\begin{align}
  \label{eq:Sudakov}
  k & = k^+ \frac{n^-}{2} + k^- \frac{n^+}{2} + {\bm k}, & k^\pm & = n^\pm\cdot k.
\end{align}
High energy factorized amplitudes reveal strong ordering in  plus and minus components of momenta which is reflected in the following kinematic constraint obeyed by the 
reggeized gluon field 
\begin{align}
  \label{eq:kinematic}
  \partial_+ A_- (x)& = 0 = \partial_+ A_+(x).
\end{align}
Even though the reggeized gluon field is charged under the QCD gauge
group SU$(N_c)$, it is invariant under local gauge transformations:
$\delta A_\pm = 0$.  Its kinetic term and the gauge invariant coupling
to the QCD gluon field are contained in the induced term,
\begin{align}
\label{eq:1efflagrangian}
  S_{\text{ind.}} = \int \text{d}^4 x \,
\text{tr}\left[\left(W_-[v(x)] - A_-(x) \right)\partial^2_\perp A_+(x)\right]
+\text{tr}\left[\left(W_+[v(x)] - A_+(x) \right)\partial^2_\perp A_-(x)\right],
\end{align}
with 
\begin{align}
  \label{eq:funct_expand}
  W_\pm[v(x)] =&
v_\pm(x) \frac{1}{ D_\pm}\partial_\pm
,
&
D_\pm & = \partial_\pm + g v_\pm (x).
\end{align}
For a more in depth discussion of the effective action we refer the reader to \cite{LevSeff} and  the recent review \cite{review}.

\subsection{Feynman rules and regularization}
\label{sec:frandreg}

\begin{figure}[htb]
    \label{fig:subfigures}
   \centering
   \parbox{.7cm}{\includegraphics[height = 1.8cm]{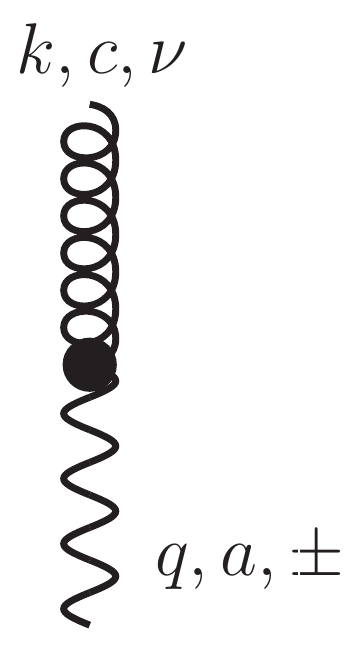}} $=  \displaystyle 
   \begin{array}[h]{ll}
    \\  \\ - i{\bm q}^2 \delta^{a c} (n^\pm)^\nu,  \\ \\  \qquad   k^\pm = 0.
   \end{array}  $ 
 \parbox{1.2cm}{ \includegraphics[height = 1.8cm]{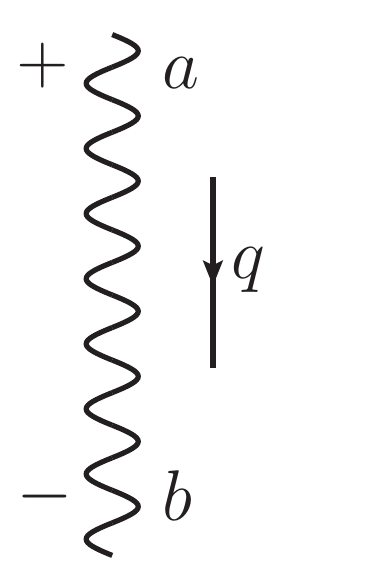}}  $=  \displaystyle    \begin{array}[h]{ll}
    \delta^{ab} \frac{ i/2}{{\bm q}^2} \end{array}$ 
 \parbox{1.7cm}{\includegraphics[height = 1.8cm]{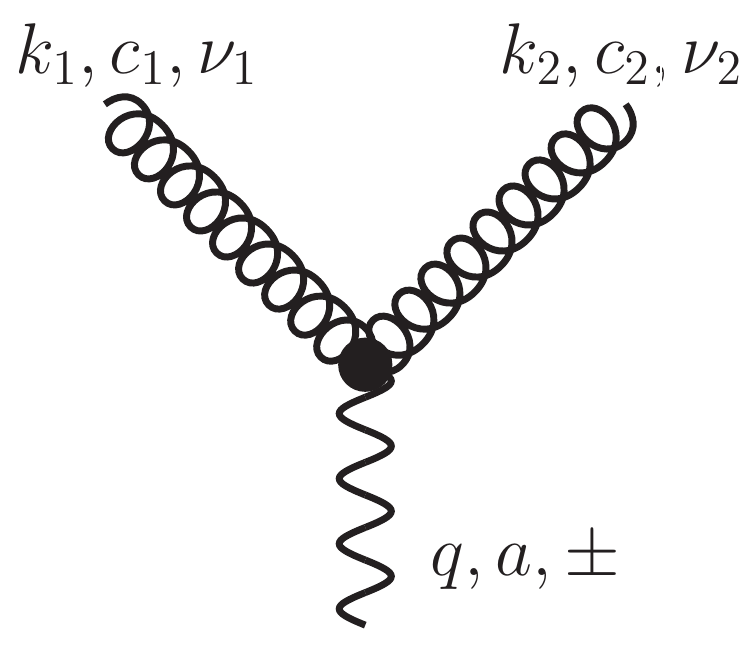}} $ \displaystyle  =  \begin{array}[h]{ll}  \\ \\ g f^{c_1 c_2 a} \frac{{\bm q}^2}{k_1^\pm}   (n^\pm)^{\nu_1} (n^\pm)^{\nu_2},  \\ \\ \quad  k_1^\pm  + k_2^\pm  = 0.
 \end{array}$
 \\
\parbox{4cm}{\center (a)} \parbox{4cm}{\center (b)} \parbox{4cm}{\center (c)}

\vspace{1cm}
  \parbox{2.4cm}{\includegraphics[height = 1.8cm]{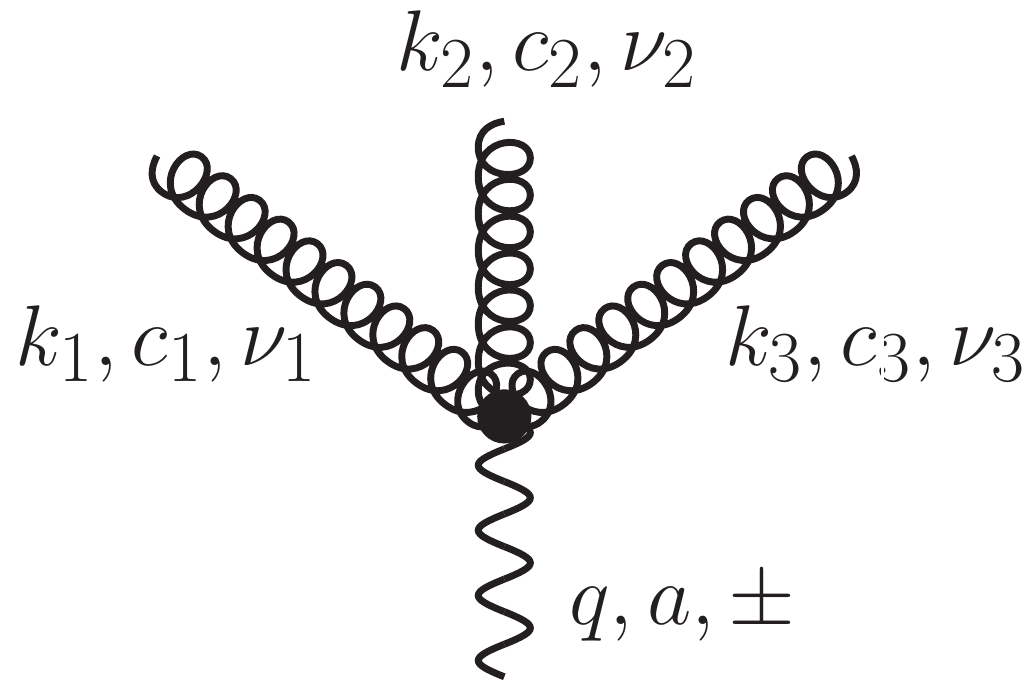}} $ \displaystyle 
   \begin{array}[h]{l}  \displaystyle  \\ \displaystyle= ig^2 {\bm{q}}^2 
\left(\frac{f^{a_3a_2 e} f^{a_1ea}}{k_3^\pm k_1^\pm} 
+
 \frac{f^{a_3a_1 e} f^{a_2ea}}{k_3^\pm k_2^\pm}\right) (n^\pm)^{\nu_1} (n^\pm)^{\nu_2} (n^\pm)^{\nu_3}, \\ \\
\qquad \qquad   k_1^\pm + k_2^\pm + k_3^\pm = 0.
   \end{array}
$ \\ 
\vspace{.3cm}
\parbox{1cm}{(d)}

 \caption{\small Feynman rules for the lowest-order effective vertices of the effective action. Wavy lines denote reggeized fields and curly lines gluons. }
\label{fig:3}
\end{figure}

\begin{figure}[htb]
  \centering
   \parbox{3cm}{\includegraphics[height = 2cm]{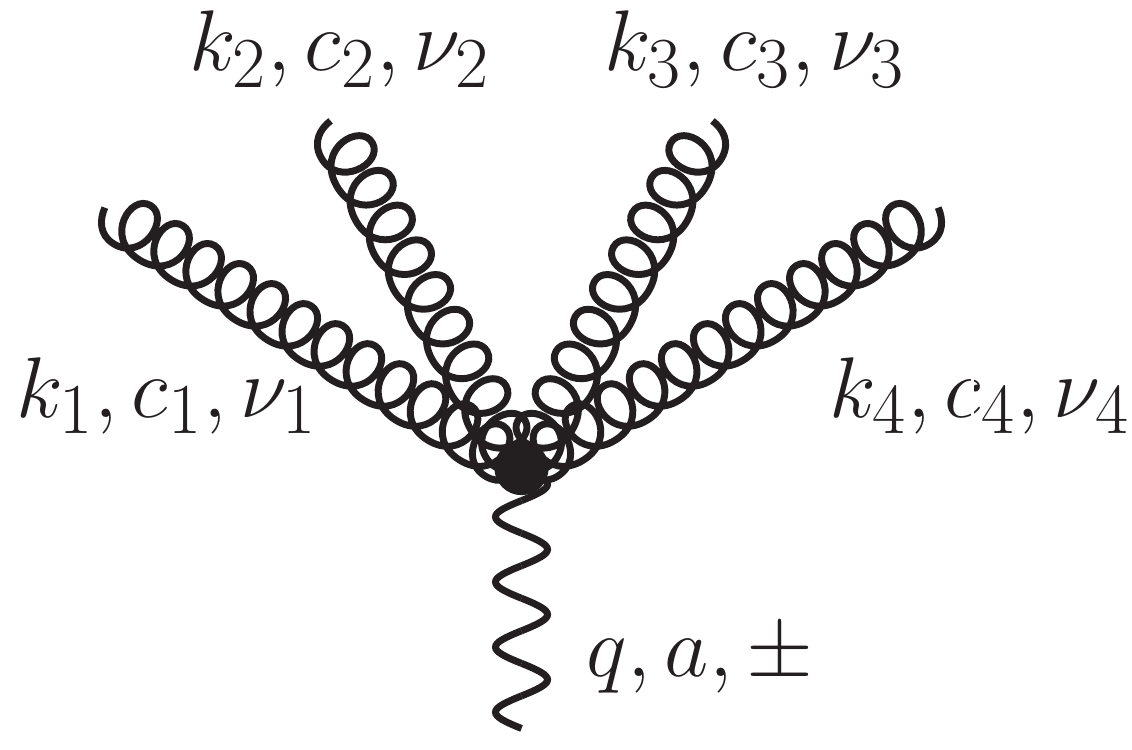}}  $ \begin{array}[h]{ll}  \\  =
   \displaystyle 
&   \displaystyle   g^3 {\bm{q}}^2  
\bigg[
 \frac{f^{c_4c_3 e_2}}{k_4^\pm}
\bigg(
\frac{f^{e_2c_1e_1} f^{c_2e_1a}}{(k_1^\pm + k_2^\pm)k_2^\pm} +  \frac{f^{e_2c_2e_1} f^{c_1e_1a}}{(k_1^\pm + k_2^\pm)k_1^\pm}
\bigg)  \\ \\ \displaystyle 
& \displaystyle  + \,
\frac{f^{c_4c_1 e_2}}{k_4^\pm}
\bigg(
\frac{f^{e_2a_2e_1} f^{c_3e_1a}}{(k_3^\pm + k_2^\pm)k_3^\pm} +
 \frac{f^{e_2a_3e_1} f^{c_2e_1a}}{(k_3^\pm + k_2^\pm)k_2^\pm}
\bigg)  \, + \\ \\
   \end{array} $ \\
$
\begin{array}[h]{l}
  \displaystyle + 
\frac{f^{c_4c_2 e_2}}{k_4^\pm}
\bigg(
\frac{f^{e_2c_1e_1} f^{c_3e_1a}}{(k_3^\pm + k_1^\pm)k_3^\pm} +  \frac{f^{e_2c_3e_1} f^{c_1e_1a}}{(k_3^\pm + k_1^\pm)k_1^\pm}
\bigg)
\bigg](n^\pm)^{\nu_1} (n^\pm)^{\nu_2}  (n^\pm)^{\nu_3}  (n^\pm)^{\nu_4},  \\ \\ \qquad \qquad  \qquad \qquad \qquad \qquad  \qquad k_1^\pm  + k_2^\pm +  k_3^\pm  + k_4^\pm = 0.
\end{array}
$
  \caption{\small The  order $g^3$ induced vertex. }
  \label{fig:trans3}
\end{figure}

Apart from the usual QCD Feynman rules, the Feynman rules of the
effective action comprise the propagator of the reggeized gluon
and an infinite number of so-called induced vertices, which result
from the non-local functional Eq.~\eqref{eq:funct_expand}.  Vertices
and propagators needed for the current study are collected in
Fig.~\ref{fig:3} and Fig.~\ref{fig:trans3}.

Loop diagrams of the effective action lead to a new type of
longitudinal divergences which are not present in conventional quantum
corrections to QCD amplitudes, and can be regularized introducing an external parameter $\rho$, evaluated in the limit $\rho \to \infty$, which deforms the light-like vectors $n^\pm$ into
\begin{align}
  \label{eq:deform}
  n^-  & \to n_a = e^{-\rho} n^+ + n^-, \notag \\
  n^+ & \to n_b = n^+ + e^{-\rho} n^-,
\end{align}
without violating the gauge invariance properties of the induced term
Eq.~\eqref{eq:1efflagrangian}.  While it is possible to identify $\rho$
with a logarithm in  $s$ or the rapidity interval spanned by a certain high energy
process, we refrain from such an interpretation and consider in the
following $\rho$ as an external parameter, similar to the parameter
$\epsilon$ in dimensional regularization in $d = 4 + 2 \epsilon$
dimensions.

\subsection{Pole prescription}
\label{sec:pole}

The evaluation of loop diagrams requires a prescription to
circumvent the light-cone singularities in the induced vertices shown in 
Figs.~\ref{fig:3},\ref{fig:trans3}.  The seemingly natural choice
which is to simply replace the operator $D_\pm$ in
Eq.~\eqref{eq:funct_expand} by {\it e.g.} $D_\pm - \epsilon$ does not
work in this context as it spoils hermiticity of the effective action.
At the level of Feynman diagrams this is reflected by terms that violate high energy
factorization. Both effects can be traced back to the existance of
new symmetric color tensors, not present in the vertices of
Figs.~\ref{fig:3},~\ref{fig:trans3}. For a more in depth discussion we
refer to \cite{Hentschinski:2011xg}. This problem can be solved by
systematically projecting out these symmetric color structures, order
by order in perturbation theory, sticking in this way to the
color tensors present in the original vertices
Fig.~\ref{fig:3},~\ref{fig:trans3}. The resulting pole prescription
respects then Bose symmetry of the induced vertices and high energy
factorization \cite{Hentschinski:2011xg}. The $\mathcal{O}(g)$ vertex is taken as a  Cauchy principal value:
\begin{align}
  \label{eq:indu1eps}
   \parbox{2.2cm}{\includegraphics[height = 2cm]{indu1.pdf}} &  = g f^{c_1 c_2 a}   \frac{{\bm q}^2}{[k_1^\pm]}   (n^\pm)^{\nu_1} (n^\pm)^{\nu_2}, & \frac{1}{[k_1^\pm]} & \equiv   \frac{1}{2} \left( \frac{1}{k_1^\pm + i\epsilon }  + \frac{1}{k_1^\pm - i\epsilon}  \right).
\end{align}
For the $\mathcal{O}(g^2)$ and $\mathcal{O}(g^3)$ vertices the light-cone denominators are to  be replaced by certain functions\footnote{We corrected a  typing error present in Eq.~(16) of \cite{Hentschinski:2011xg} in the expression below.} $g_2$ and $g_3$:
\begin{align}
  \label{eq:double_comm_simpli}
   \parbox{2.6cm}{\includegraphics[height = 2cm]{indu2.pdf}}
   \begin{array}[h]{l } \\
\displaystyle
 =  -ig^2{\bm q}^2    \bigg[ f^{c_3c_2e}f^{c_1ea}
   g_2^\pm(3,2,1) \\  \displaystyle  \qquad \qquad \qquad 
          +  f^{c_3c_1e}f^{c_2ea}
       g_2^\pm(3,1,2)\bigg] n^\pm_{\nu_1} n^\pm_{\nu_2} n^\pm_{\nu_3} ,     
   \end{array}
\end{align}
\begin{align}
  \label{eq:pole_indu3}
& \parbox{3cm}{\includegraphics[height = 2cm]{indu3.pdf}}  = 
-g^3 {\bm q}^2
n^\pm_{\nu_1} n^\pm_{\nu_2} n^\pm_{\nu_3} n^\pm_{\nu_4} \, \cdot  \notag \\
\bigg[&
 f^{a_4a_1d_2}f^{d_2a_3d_1}f^{d_1a_2c} 
 g_3^\pm(4,1,3,2)
+
f^{a_4a_1d_2}f^{d_2a_2d_1}f^{d_1a_3c}
 g_3^\pm(4,1,2,3)
\notag \\
+&
f^{a_4a_2d_2}f^{d_2a_1d_1}f^{d_1a_3c}
 g_3^\pm(4,2,1,3)
+
f^{a_4a_2d_2}f^{d_2a_3d_1}f^{d_1a_1c}
 g_3^\pm(4,2,3,1)
\notag \\
+&
f^{a_4a_3d_2}f^{d_2a_1d_1}f^{d_1a_2c}
 g_3^\pm(4,3,1,2)
+
f^{a_4a_3d_2}f^{d_2a_2d_1}f^{d_1a_1c}
 g^\pm_3(4,3,2,1) 
\bigg].
\end{align}
They are obtained as 
\begin{align}
  \label{eq:cpv_rep}
g_2^\pm(i,j,m) = 
    \bigg[&  \frac{-1}{[k_i^\pm][k_m^\pm]} -\frac{\pi^2}{3}\delta(k_i^\pm)\delta(k_m^\pm) \bigg].
\end{align}
and 
\begin{align}
  \label{eq:f_cpv}
g_3^\pm(i,j,m,n)
 =\bigg(&
          \frac{-1}{[k_i^\pm][k_n^\pm + k_m^\pm][k_n^\pm]}
            -
            \frac{\pi^2}{3} \delta(k_n^\pm) \delta(k_m^\pm) \frac{-1}{[k_i^\pm]} 
             \notag \\
             &
-
            \frac{\pi^2}{3} \delta(k_n^\pm) \delta(k_i^\pm) \frac{1}{[ k_m^\pm]} 
        - 
          \frac{\pi^2}{3} \delta(k_n^\pm + k_m^\pm) \delta(k_i^\pm) \frac{1}{[k_n^\pm]} \bigg).
\end{align}

\section{The gluon Regge trajectory from  the effective action}
\label{sec:gluontrajectory-effectiveaction}

A key  ingredient in  the resummation of high energy logarithms of QCD
scattering amplitudes is provided by a universal function associated
with the exchange of a single reggeized gluon, known as the Regge
trajectory of the gluon.  
For the  real part of  QCD scattering amplitudes, where the high energy description is given in terms of single reggeized gluon exchange, this function is known to govern the entire energy dependence at leading logarithmic (LL) and
next-to-leading logarithmic (NLL) accuracy.

Multiple reggeized gluon exchanges appear on the other hand for the
high energy description of the imaginary part of scattering amplitudes
and in general for amplitudes beyond NLL accuracy. While this requires
new elements, which describe in a nutshell the interaction between
reggeized gluons, the gluon Regge trajectory remains an essential
building block in the formulation of high energy resummation also in
this more general case.

\begin{figure}[htb]
  \centering
  \parbox{1.5cm}{\vspace{0.1cm} \includegraphics[height = 2.5cm]{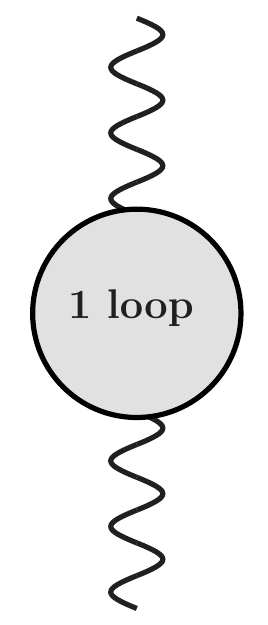}}
= 
    \parbox{1cm}{\vspace{0.1cm} \includegraphics[height = 2.5cm]{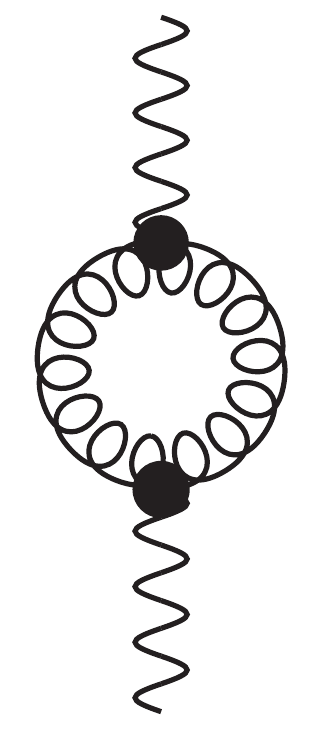}}
  + 
  \parbox{1cm}{\vspace{0.1cm} \includegraphics[height = 2.5cm]{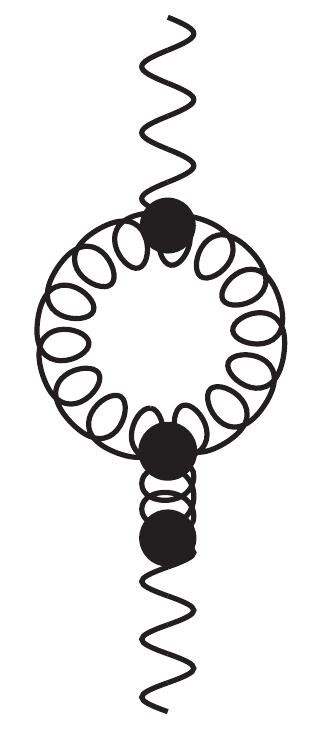}} 
 +
  \parbox{1cm}{\vspace{0.1cm} \includegraphics[height = 2.5cm]{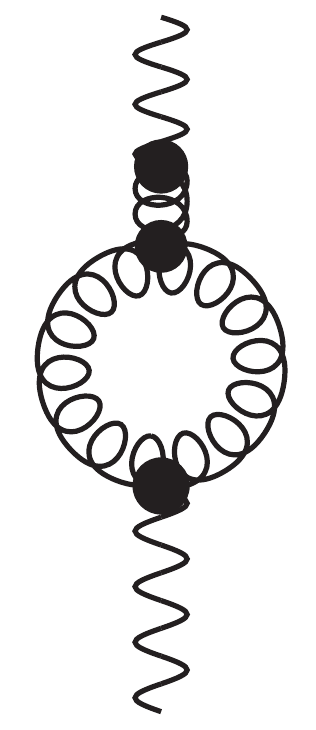}} 
 +
  \parbox{1cm}{\vspace{0.1cm} \includegraphics[height = 2.5cm]{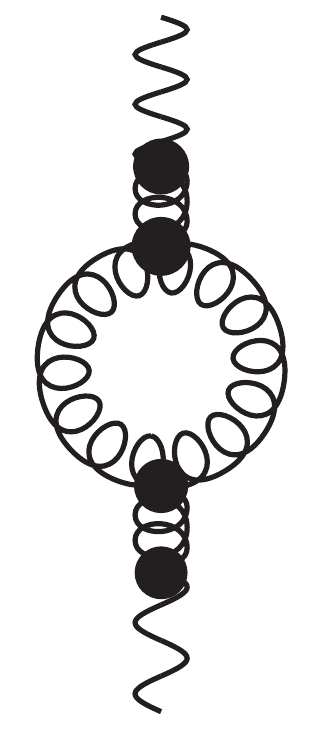}} 
 +
  \parbox{1cm}{\vspace{0.1cm} \includegraphics[height = 2.5cm]{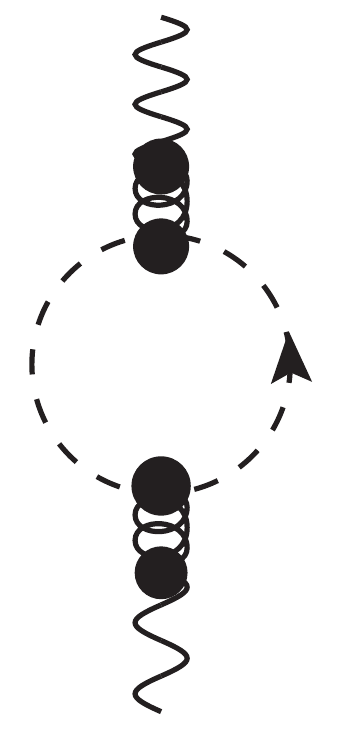}}
+
  \parbox{1cm}{\vspace{0.1cm} \includegraphics[height = 2.5cm]{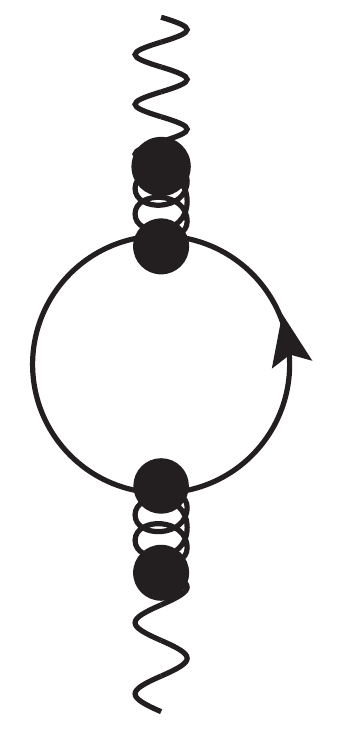}}
\caption{\small Diagrams contributing to the one-loop reggeized gluon self-energy.}
\label{fig:self_1loop}
\end{figure}

To be more precise, for the elastic process $p_a + p_b \to p_1 + p_2$ with $s = (p_a + p_b)^2$ and $t = q^2$ with $q = p_a - p_1$ one finds for  amplitudes with gluon quantum numbers in the $t$-channel  at LL and NLL accuracy  the following factorized form\footnote{For a pedagogical review see \cite{Fadin:1998sh}.}
\begin{align}
  \label{eq:M8}
 \frac{ \mathcal{M}_{(\bf 8_A)} (s,t)}{ \mathcal{M}^{(0)} (s,t) } & = \Gamma_{a1}(t)  \left[ \left(\frac{-s}{-t} \right)^{\omega(t)}  +   \left(\frac{s}{-t} \right)^{\omega(t)} \right]  \Gamma_{b2}(t), 
\end{align}
where $\mathcal{M}^{(0)}_{(\bf 8_A)}$ is the tree-level amplitude
and the subscript `${\bf 8_A}$' denotes that the allowed $t$-channel
exchange is restricted to the anti-symmetric color octet channel.
The functions $\Gamma_{ij}(t)$ are known as impact factors, describing
the coupling of the reggeized gluons to scattering particles. For the
case of gluon and quarks they have been determined within the
effective action in \cite{quarkjet, gluonjet}. The function
$\omega(t)$ which governs the $s$-dependence of the scattering
amplitude is on the other hand the Regge trajectory of the gluon.  It
is currently known to leading \cite{BFKL1} and next-leading order
\cite{Fadin:1996tb} for QCD and to all orders in $\mathcal{N}=4$ super
Yang-Mills theory \cite{Bartels:2008ce}.
The procedure which allows the  derivation of the gluon trajectory from
the effective action has  been originally discussed in \cite{Chachamis:2012gh}.  It consists of two steps
\begin{itemize}
\item determination of the propagator of the reggeized gluon to the desired order in $\alpha_s$;
\item renormalization of the rapidity divergences of the reggeized
  gluon propagator;  the gluon Regge trajectory is then identified as the
  coefficient of the $\rho$ dependent term in the renormalization
  factor.
\end{itemize}
 To obtain the reggeized gluon propagator to order $\alpha_s^2$ it is needed to determine the one- and two-loop self-energies of the reggeized gluon. Following the subtraction procedure proposed in \cite{quarkjet} these self-energies can be obtained through 
\begin{itemize}
\item
 determination of the self-energy of the reggeized gluon from the effective action, with the reggeized gluon treated as a background field;
\item
subtraction of all  disconnected contributions which contain internal reggeized gluon lines.
\end{itemize}
Using a symmetric pole prescription as given in Sec.~\ref{sec:pole},
all  diagrams with internal reggeized gluon lines that would possibly contribute to the one
loop self energy can be shown to vanish and no subtraction is
necessary.  The contributing diagrams are shown in
Fig.~\ref{fig:self_1loop}.
\\
Keeping the ${\cal O} (\rho, \rho^2)$, for $\rho  \to \infty$, terms and using the notation
\begin{align}
  \label{eq:gbar}
  \bar{g}^2 & =  \frac{g^2 N_c \Gamma(1 - \epsilon)}{(4 \pi)^{2 + \epsilon}},
\end{align}
we have the following result in $d = 4 + 2\epsilon$ dimensions\footnote{In the original result presented in \cite{quarkjet} and reproduced in \cite{Chachamis:2012gh,review}  a  finite result for the second and third diagram has been erroneously included. This has been corrected in the result presented here.}:
\begin{align}
\label{eq:self_1loop}
 &  \parbox{1cm}{\includegraphics[width = 1cm]{self_1loop.pdf}} 
=    \Sigma^{(1)}\left(\rho; \epsilon, \frac{{\bm q}^2}{\mu^2}    \right)      
\notag \\
& \hspace{.4cm}= 
 \frac{(-2i {\bm q}^2) \bar{g}^2 \Gamma^2(1 + \epsilon)}{\Gamma(1 + 2 \epsilon)} \left(\frac{{\bm q}^2}{\mu^2} \right)^\epsilon  
  \bigg\{  \frac{ i\pi - 2 \rho}{\epsilon}         
- \frac{1}{(1 + 2 \epsilon)\epsilon} \bigg[   \frac{5 + 3\epsilon}{3 + 2 \epsilon} 
-\frac{n_f}{N_c}  \left(\frac{2 + 2\epsilon}{3 + 2\epsilon}\right)\bigg] \bigg\}. 
\end{align}
To determine the 2-loop self energy it is on the other hand needed to subtract disconnected diagrams, whereas diagrams with multiple internal reggeized gluons can be shown to yield a zero result, if the symmetric pole prescription of Sec.~\ref{sec:pole} is used. Schematically one has 
\begin{align}
  \label{eq:coeff_2loop}
  \Sigma^{(2)}\left(\rho; \epsilon, \frac{{\bm q}^2}{\mu^2}    \right)   & =   \parbox{2cm}{\center \includegraphics[height = 2.5cm]{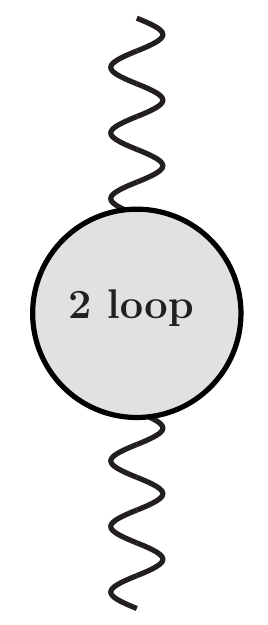}}  
  = 
  \parbox{2cm}{\center \includegraphics[height = 2.5cm]{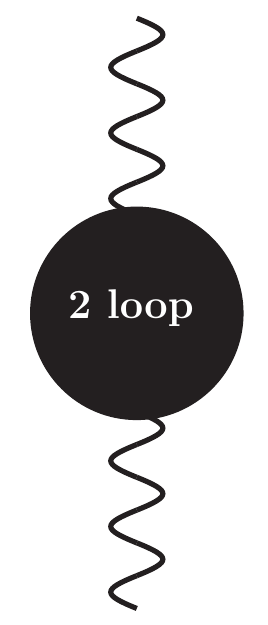}}
  -
  \parbox{2cm}{\center \includegraphics[height = 2.5cm]{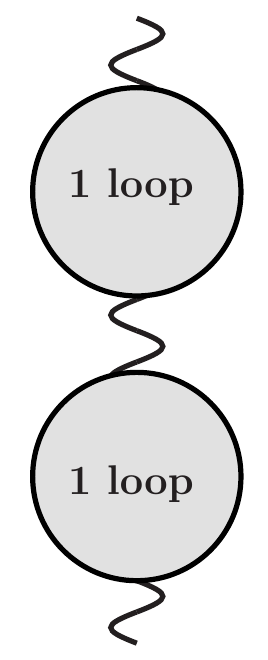}},
\end{align}
where the black blob denotes the unsubtracted 2-loop reggeized gluon
self-energy, which is obtained through the direct application of the
Feynman rules of the effective action, with the reggeized gluon itself
treated as a background field. Its determination will be discussed in
detail in the forthcoming section.  The (bare) two-loop reggeized
gluon propagators then reads
\begin{align}
  \label{eq:barepropR}
   G \left(\rho; \epsilon, {\bm q}^2, \mu^2   \right)
&=
\frac{i/2}{{\bm q}^2} \left\{ 1 + \frac{i/2}{{\bm q}^2} \Sigma \left(\rho; \epsilon, \frac{{\bm q}^2}{\mu^2}    \right)  + \left[  \frac{i/2}{{\bm q}^2} \Sigma \left(\rho; \epsilon, \frac{{\bm q}^2}{\mu^2}   \right)\right] ^2 + \ldots   \right\},
\end{align}
with
\begin{align}
  \label{eq:sigma}
   \Sigma \left(\rho; \epsilon, \frac{{\bm q}^2}{\mu^2}   \right) & =   \Sigma^{(1)} \left(\rho; \epsilon, \frac{{\bm q}^2}{\mu^2}   \right) +  \Sigma^{(2)} \left(\rho; \epsilon, \frac{{\bm q}^2}{\mu^2}   \right) + \ldots
\end{align}
where the dots indicate higher order terms. As discussed in
Sec.~\ref{sec:frandreg} and as directly apparent from
Eq.~\eqref{eq:self_1loop}, the reggeized gluon self-energies are
divergent in the limit $\rho \to \infty$. In \cite{quarkjet, gluonjet}
it has been demonstrated by explicit calculations that these
divergences cancel at one-loop level, for both quark-quark and
gluon-gluon scattering amplitudes, against divergences in the
couplings of the reggeized gluon to external particles. The entire
one-loop amplitude is then found to be free of any high energy
singularity in $\rho$.  High energy factorization then suggests that
such a cancellation holds also beyond one loop. Starting from this assumption, it is possible to define a
renormalized reggeized gluon propagator through
\begin{align}
\label{eq:renor}
G^{\rm R}(M^+,M^-;\epsilon,\bm{q}^2,\mu^2)=\frac{G(\rho;\epsilon,\bm{q}^2,\mu^2)}{Z^+\left(\frac{M^+}{\sqrt{\bm{q}^2}},\rho;\epsilon,\frac{\bm{q}^2}{\mu^2}\right)Z^-\left(\frac{M^-}{\sqrt{\bm{q}^2}},\rho;\epsilon,\frac{\bm{q}^2}{\mu^2}\right)},
\end{align}
where the renormalization factors need to cancel against corresponding
renormalization factors associated with the vertex to which the
reggeized gluon couples with `plus' ($Z^+$) and `minus' ($Z^-$)
polarization. For explicit examples we refer the reader to
\cite{gluonjet,Chachamis:2012gh}.  In their most general form these
renormalization factors are parametrized as
\begin{align}
\label{eq:param}
Z^\pm\left(\frac{M^\pm}{\sqrt{\bm{q}^2}},\rho;\epsilon,\frac{\bm{q}^2}{\mu^2}\right)=\exp\left[\left(\frac{\rho}{2}-\ln\frac{M^\pm}{\sqrt{\bm{q}^2}}\right)\omega\left(\epsilon,\frac{\bm{q}^2}{\mu^2}\right)   +   f^\pm\left(\epsilon,\frac{\bm{q}^2}{\mu^2}\right)\right],
\end{align}
 with the coefficient of the $\rho$-divergent term  given by the gluon Regge trajectory $\omega(\epsilon, {\bm q}^2)$. 
It is assumed to have the  following perturbative expansion
\begin{align}
  \label{eq:perturexpomega}
  \omega\left(\epsilon, \frac{{\bm q}^2}{\mu^2} \right) &= 
 \omega^{(1)}\left(\epsilon, \frac{{\bm q}^2}{\mu^2} \right)  
+
 \omega^{(2)}\left(\epsilon, \frac{{\bm q}^2}{\mu^2} \right)  + \ldots,
\end{align}
and is to be determined by the requirement that the renormalized reggeized
gluon propagator must, at each loop order, be free of  $\rho$
divergences.  At one loop we get from Eq.~\eqref{eq:self_1loop}
\begin{align}
  \label{eq:omega1}
  \omega^{(1)}\left(\epsilon, \frac{{\bm q}^2}{\mu^2} \right) & = -\frac{2 \bar{g}^2 \Gamma^2(1 + \epsilon)}{\Gamma(1 + 2 \epsilon)\epsilon } \left(\frac{{\bm q}^2}{\mu^2} \right)^\epsilon.  
\end{align}
The function $f^\pm(\epsilon, {\bm q}^2)$ parametrizes finite
contributions and is, in principle, arbitrary. While symmetry of the
scattering amplitude requires $f^+ = f^- = f$, Regge theory suggests
fixing it in such a way that terms which are not enhanced in $\rho$
are entirely transferred from the reggeized gluon propagators to the vertices, to which the reggeized gluon
couples. With the perturbative expansion
\begin{align}
  \label{eq:expansionofF}
  f\left(\epsilon, \frac{{\bm q}^2}{\mu^2} \right) & = f^{(1)}\left(\epsilon, \frac{{\bm q}^2}{\mu^2} \right) + f^{(2)}\left(\epsilon, \frac{{\bm q}^2}{\mu^2} \right) \ldots
\end{align}
we obtain from  Eq.~\eqref{eq:self_1loop}
\begin{align}
  \label{eq:f1loop}
   f^{(1)}\left(\epsilon, \frac{{\bm q}^2}{\mu^2} \right) & =  \frac{ \bar{g}^2 \Gamma^2(1 + \epsilon)}{\Gamma(1 + 2 \epsilon)} \left(\frac{{\bm q}^2}{\mu^2} \right)^\epsilon  
         \frac{(-1)}{(1 + 2 \epsilon)2 \epsilon} \bigg[    \frac{5 + 3\epsilon}{3 + 2 \epsilon} 
-\frac{n_f}{N_c} \left( \frac{2 + 2\epsilon}{3 + 2\epsilon} \right)\bigg]  .
\end{align}
The  renormalized  reggeized gluon propagator is then to one loop accuracy given by
\begin{align}
\label{eq:renor}
G^{\rm R}(M^+,M^-;\epsilon,\bm{q}^2,\mu^2) & = 1 +  \omega^{(1)}\left(\epsilon, \frac{{\bm q}^2}{\mu^2} \right) \left( \log \frac{M^+M^-}{{\bm q}^2} - \frac{i \pi}{2} \right) + \ldots 
\end{align}
The scales $M^+$ and $M^-$ are arbitrary; their role is analogous to
the renormalization scale in UV renormalization and the factorization
scale in collinear factorization. They are naturally chosen to
coincide with the corresponding light-cone momenta of scattering
particles to which the reggeized gluon couples. To determine the gluon Regge trajectory at two loops we need in addition the $\rho$-enhanced terms of the two-loop reggeized gluon self-energy. From Eq.~\eqref{eq:renor} we  obtain the following relation
\begin{align}
  \label{eq:omega2_defined}
 \omega^{(2)}\left(\epsilon, \frac{{\bm q}^2}{\mu^2} \right)  & = 
 \lim_{\rho \to \infty} \frac{1}{\rho} \bigg[ 
\frac{ \Sigma^{(2)}}{(-2 i {\bm q}^2)} 
+
\left( \frac{ \Sigma^{(1)}} {(-2 i {\bm q}^2)}  \right)^2
-
     \left(\rho  \omega^{(1)}  + 2  f^{(1)}  \right) \frac{\Sigma^{(1)}} {(-2 i {\bm q}^2)} \notag \\
& \qquad \qquad \qquad \qquad \qquad \qquad  \qquad \qquad
+ \frac{\rho^2}{2} \left(\omega^{(1)}\right)^2 + 2 \rho   f^{(1)} \omega^{(1)} \bigg]
\notag \\
& =
 \lim_{\rho \to \infty} \frac{1}{\rho} \bigg[ \frac{\Sigma^{(2)}}{(-2i{\bm q}^2)}  + \frac{\rho^2}{2} \left(\omega^{(1)}\right)^2 + 2 \rho   f^{(1)} \omega^{(1)}  \bigg],
\end{align}
where we omitted at the right hand side the dependencies on $\epsilon$
and ${\bm q}^2/\mu^2$; in the last line we further expanded
$\Sigma^{(1)}$ in terms of the functions $\omega^{(1)}$ and
$f^{(1)}$. We stress that this is a non-trivial definition and that it
is not clear a priori whether the right hand side even exists due to
the presence of the second term, linear in $\rho$.  Confirmation of
this relation provides therefore an important non-trivial check on the
validity of our formalism.

\section{Computation of the 2-Loop reggeized gluon self-energy}
\label{sec:computation-two-loop}

\begin{figure}[htp]
\centering
\includegraphics[width = .9\textwidth]{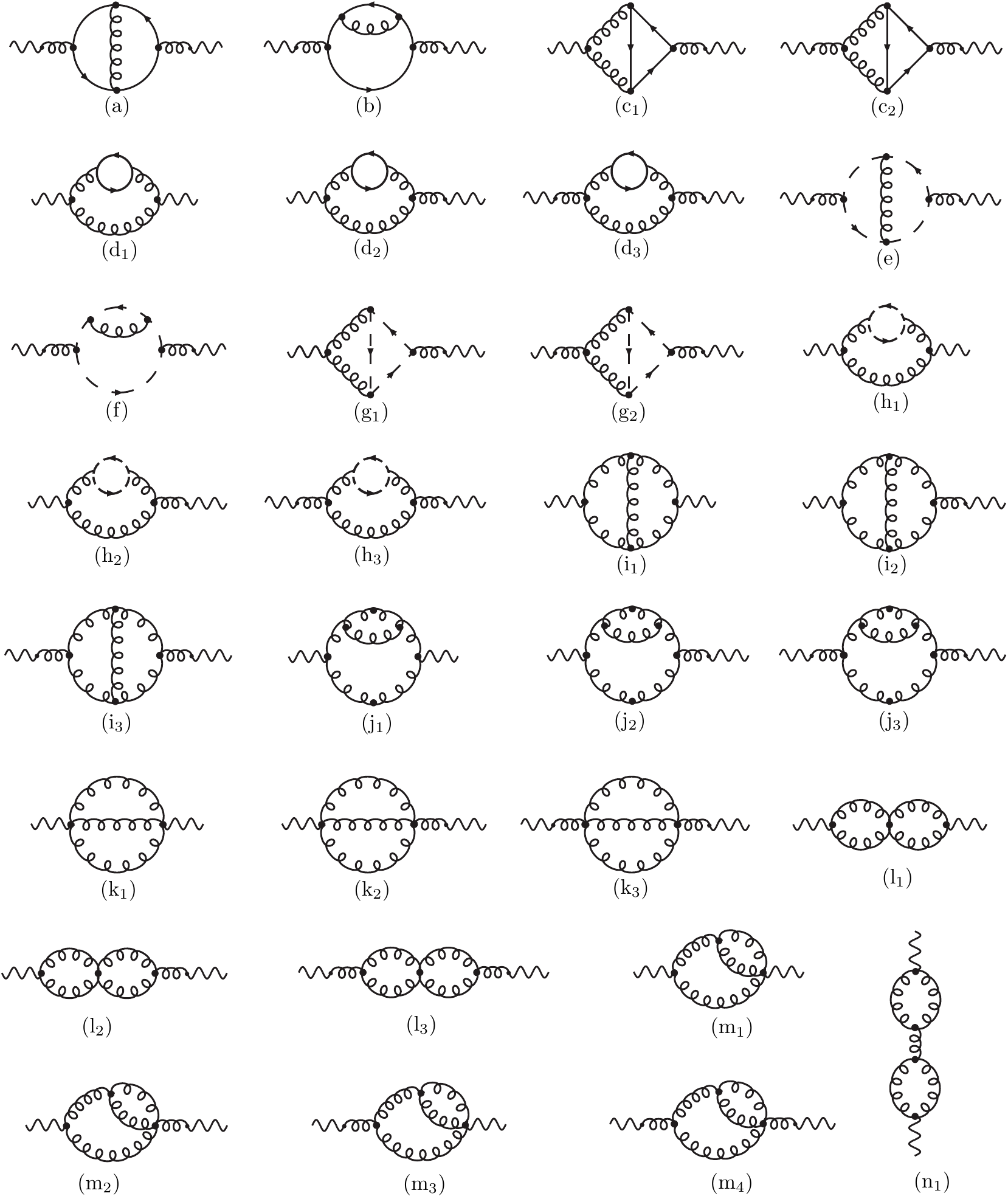}
\caption{Diagrams for the two-loop trajectory in the effective action formalism. Tadpole-like contributions are zero in dimensional regularization and are omitted.}
\label{diagram}
\end{figure}

The necessary  diagrams for the computation of the unsubtracted reggeized gluon self-energy are  shown in Fig.~\ref{diagram}.
 The diagrams (a$_1$)-(d$_3$),  containing internal quark loops, generating an overall factor  $n_f$, have been   computed in \cite{Chachamis:2012gh} and lead to the following result,
\begin{align}
 \parbox{1.5cm}{\vspace{0.1cm} \includegraphics[height = 2.5cm]{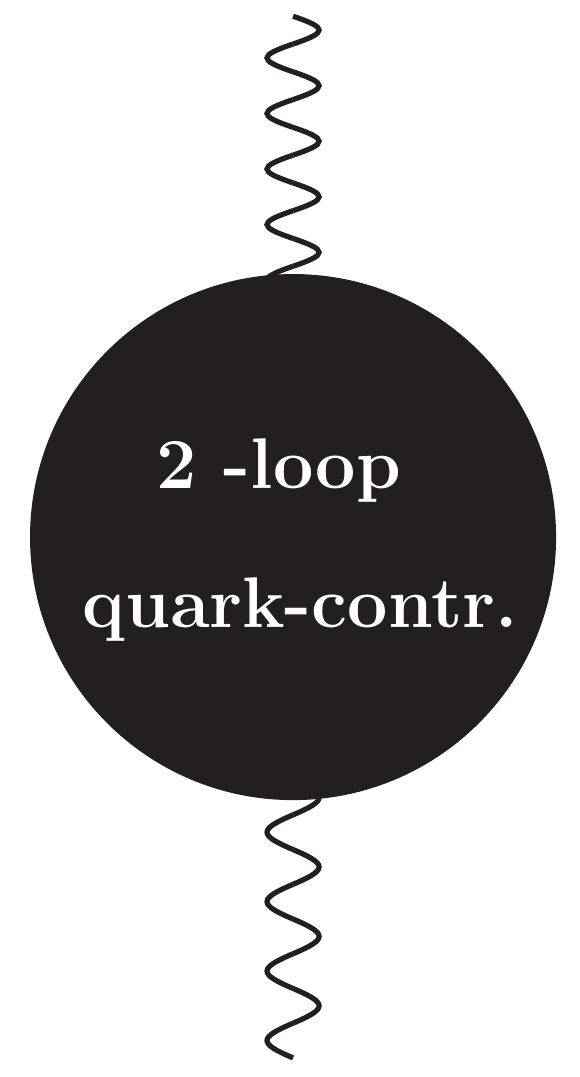}}   & =
-\rho  (-i 2 {\bm q}^2) \bar{g}^4 \frac{4 n_f}{\epsilon N_c}  \frac{\Gamma^2(2 + \epsilon)}{\Gamma(4 + 2\epsilon)} \cdot \frac{3 \Gamma(1 - 2\epsilon) \Gamma(1 + \epsilon) \Gamma(1 + 2\epsilon)}{\Gamma^2(1 - \epsilon) \Gamma(1 + 3 \epsilon) \epsilon} + \mathcal{O}(\rho^0) \, .
  \label{eq:traj_jose}
\end{align}

\subsection{The scaling argument}
\label{sec:scaling}
For the computation of the remaining diagrams we observe at first that the 
 number of diagrams, which  can be potentially enhanced by a factor $\rho^k$, $k \geq 1$, is largely reduced by scaling arguments: only those diagrams where both reggeized gluons couple to the
 internal gluon lines through induced reggeized gluon--$n$-gluon
 vertices with $n \geq 2$ have the potential to lead to an enhancement
 through a factor $\rho$. This is immediately clear for diagrams where
 both reggeized gluons couple through the reggeized gluon--$1$-gluon
 vertex Fig.~\ref{fig:3} (a) to the internal QCD lines. Those diagrams are a 
 projection of the 2-loop QCD polarization tensor onto the kinematics
 of reggeized gluons and no $\rho$ enhancement can be expected.\\

\begin{figure}[htp]
\centering
\includegraphics[width = 3cm]{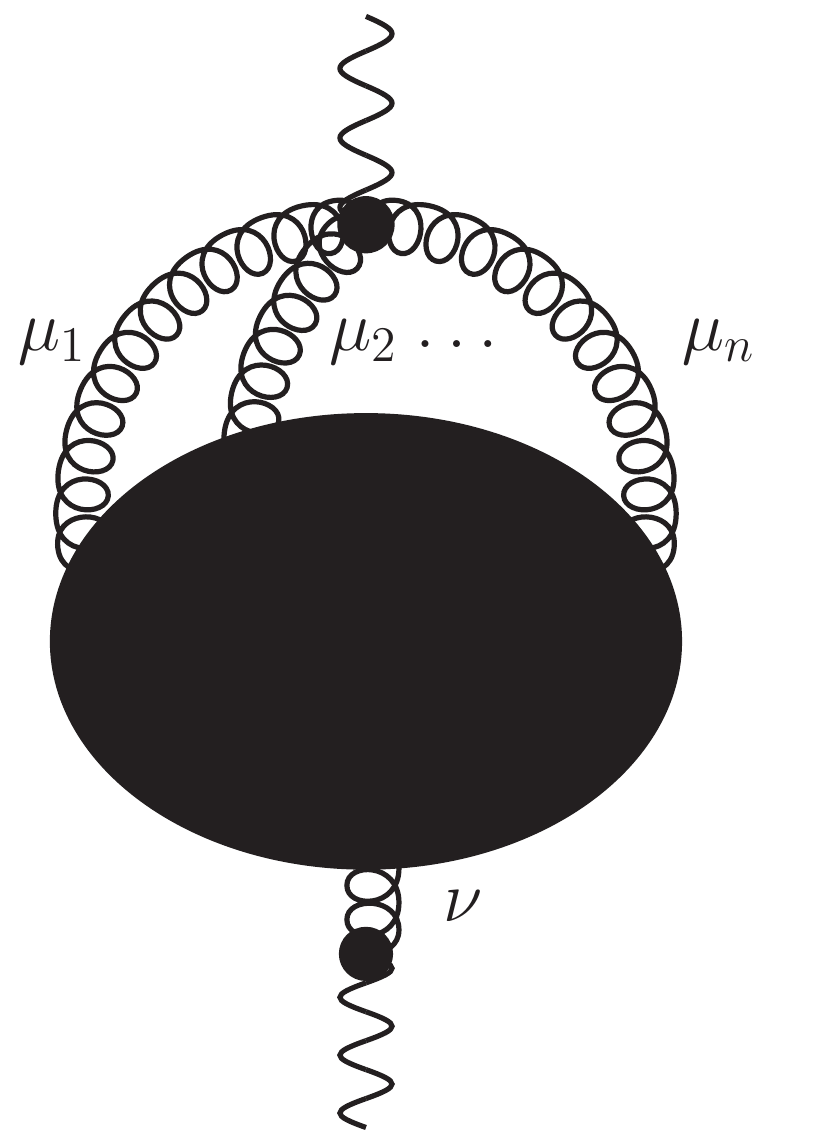}
\caption{General non-enhanced diagram.}
\label{current}
\end{figure}

To address the  case where only one of the reggeized gluons couples through an induced 
reggeized gluon--$n$-gluon ($n \geq 2$) vertex to the internal QCD particles, we consider the general diagram in  Fig.~\ref{current}.
The dependence on the light-cone vectors of the reggeized gluon--$n$-gluon
vertex in Fig.~\ref{current} is, up to permutations, of the form
$\frac{n_a^{\mu_1}n_a^{\mu_2}\cdots n_a^{\mu_d}}{n_a\cdot k_1 n_a\cdot
  k_2\cdots n_a\cdot k_{n-1}}$. The denominators $ n_a\cdot k_i$, $i =
1, \ldots n-1$ appear in the integrals that give rise to an amplitude
${\cal M}_{\mu_1\mu_2\cdots\mu_n\nu}$. In a general diagram such as
Fig.~\ref{current}, the only vectors that are not integrated over in
the amplitude are $q$, the momentum transfer, and $n_a$, which enters
through the denominators of the induced vertex. The vector $n_b$ only
contracts with the four-vector index $\nu$.  The whole diagram can be
therefore written as
\begin{equation}
n_a^{\mu_1}n_a^{\mu_2}\cdots n_a^{\mu_n}{\cal M}_{\mu_1\mu_2\cdots\mu_n\nu}(n_a,q)n_b^\nu.
\end{equation}
As a consequence, the tensor structure of ${\cal
  M}_{\mu_1\mu_2\cdots\mu_n\nu}(n_a,q)$ can only consist of
combinations of the four vector $n_a^\mu$ and the metric tensor
$g_{\mu\nu}$, since the external reggeized gluons imply   $q\cdot
n_a=q\cdot n_b=0$.  The only scalar combinations that can appear are
therefore $\bm{q}^2$ and $n_a^2$.  These factors must give the
dimensions required by scale transformations. If $s$ is the number of
metric tensors in the numerator for a given term and $l$ the number of
$n_a^\mu$ numerators, then $n+1=2s+l$ and  the associated scalar function  must scale as
\begin{align}
  \label{eq:scale1}
  \frac{1}{n_a^{n-1+l}}=\frac{1}{(n_a^2)^{d - s}}.
\end{align}
Next, we consider the contractions with the vertex currents. If
$n_b^\rho$ is contracted through a metric tensor then we obtain
\begin{align}
  \label{eq:scale2}
  (n_a^2)^l\,n_a\cdot n_b\,(n_a^2)^{s-1} & =(n_a^2)^{n-s}n_a\cdot n_b;
\end{align}
 if on the
other hand $n_b^\rho$ is directly contracted with one of the $n_a$'s,
we obtain a factor
\begin{align}
  \label{eq:scale3}
  n_a\cdot
n_b\,(n_a^2)^s(n_a^2)^{l-1} & =(n_a^2)^{n-s}n_a\cdot n_b.
\end{align}
 In both cases
the factors of $n_a^2$ cancel against corresponding factors in the
denominators and no enhancement can occur.  Thus, in our case only 
the diagrams (h$_1$), (i$_1$), (j$_1$),
(k$_1$), (l$_1$), (m$_1$) and (n$_1$) are potentially enhanced by
(powers of) $\rho$.

\begin{figure}[htb]
  \centering
  \parbox{3cm}{\center \includegraphics[height = 4cm]{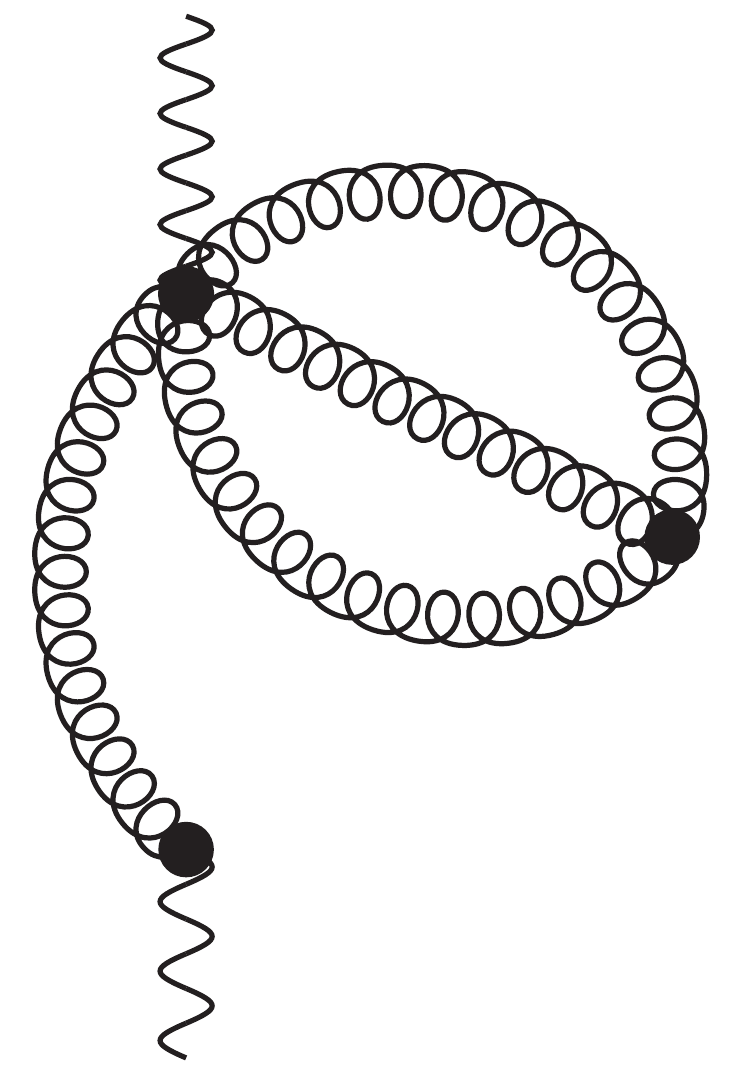}}
 \parbox{8cm}{\center  \includegraphics[height = 4cm]{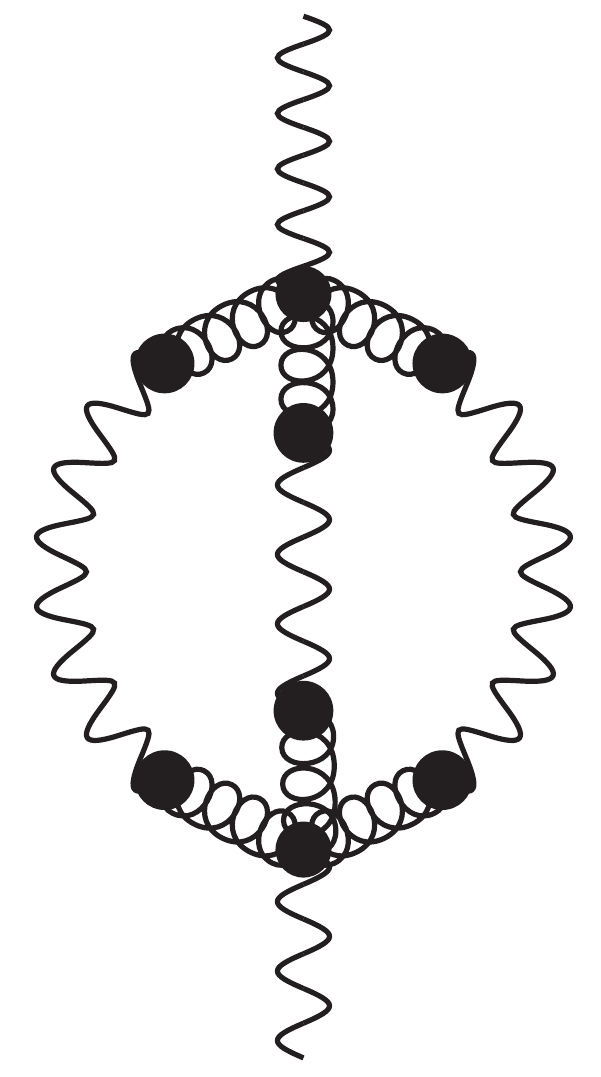}  \hspace{1cm}  \includegraphics[height = 4cm]{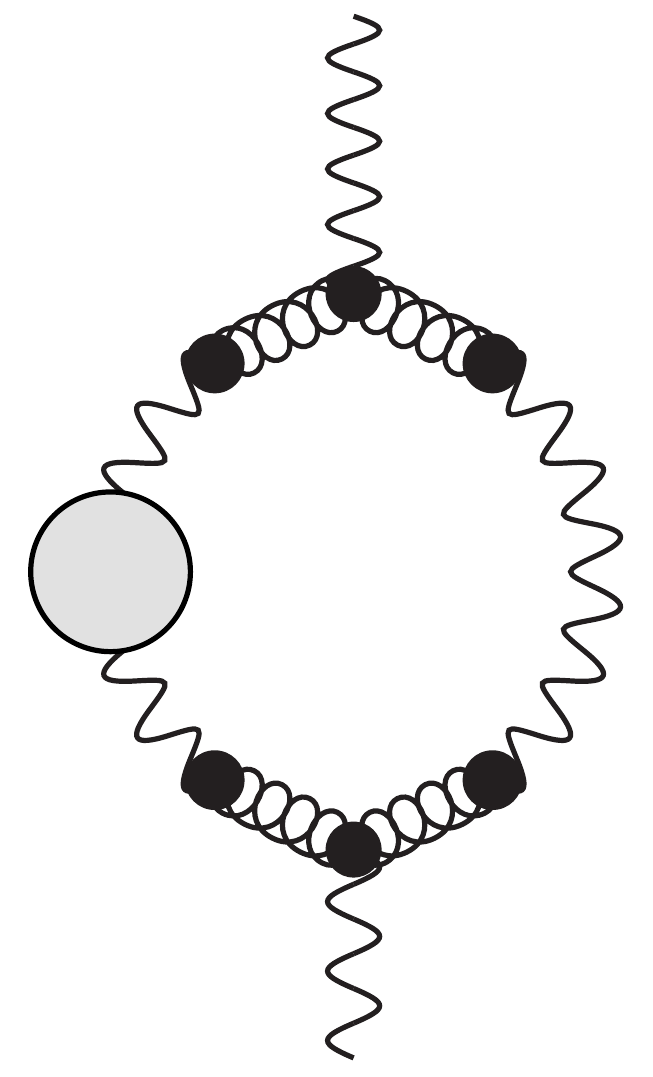}  }

\parbox{3cm}{\center (a)} \parbox{8cm}{\center (b)}
  \caption{(a)  typical tadpole contribution to the 2-loop self energy (b)  disconnected diagrams  with internal reggeized gluon loops which would contribute to  possible subtraction terms. Both contributions can be shown to vanish.}
  \label{fig:zerodiags}
\end{figure}
A further class of diagrams that can be omitted are tadpole diagrams
and diagrams with internal reggeized gluon loops. Tadpole diagrams,
such as in Fig.~\ref{fig:zerodiags} (a), have been verified to vanish in
dimensional regularization. Possible loop diagrams with internal
reggeized gluon lines, such as Fig.~\ref{fig:zerodiags} (b) vanish
identically  due to the symmetry properties obeyed by the pole
prescription of the induced vertices.

\subsection{Calculation of the enhanced diagrams}
\label{sec:calcench}

 Direct computation reveals that diagram (l$_1$) is identically zero. 
We use the notation $\xi=n_a^2=n_b^2=4e^{-\rho}$, $\delta=n_a\cdot n_b\sim 2$, and the following shorthand notation for the master integral
\begin{align}
\label{eq:MASTER}
& [\,\alpha_1, \alpha_2,  \cdots,\alpha_9\,] = 
(\mu^4)^{4-d}\iint\frac{d^dk}{(2\pi)^d}\frac{d^dl}{(2\pi)^d}\frac{1}{(-k^2-i0)^{\alpha_1}[-(k-q)^2-i0]^{\alpha_2}(-l^2-i0)^{\alpha_3} } 
\notag \\
& \times\frac{1}{
[-(l-q)^2-i0]^{\alpha_4}[-(k-l)^2-i0]^{\alpha_5}} \cdot \frac{1}{(-n_a\cdot k)^{\alpha_6}(-n_b\cdot k)^{\alpha_7}(-n_a\cdot l)^{\alpha_8}(-n_b\cdot l)^{\alpha_9}},
\end{align}
with $ n_a\cdot q=n_b\cdot q=0$ and the eikonal factors taken with the pole prescription defined in Sec.~\ref{sec:pole}. More accurately, for master integrals with single poles $1/n_{a,b}\cdot k$  ($\alpha_6 = 1, \alpha_8 = 0$, $\alpha_6 = 0, \alpha_8 = 1$ and/or  $\alpha_7 = 1, \alpha_9 = 0$, $\alpha_7 = 0, \alpha_9 = 1$)  the function $g_1^{a, b}$ is used, while  for terms with two poles   $1/n_{a,b}\cdot k/n_{a,b}\cdot l$  ($\alpha_6 = 1, \alpha_8 = 1$,  and/or  $\alpha_7 = 1, \alpha_9 = 1$)  the function  $g_2^{a,b}$ is employed.  Dropping all pieces that cannot generate terms enhanced as $\rho\to\infty$, we have the following contributions from each diagram:
\begin{equation}
\begin{aligned}
\left[i{\cal M}_{\rm h_1}\right]_{\rm enh}&=-\frac{3ig^4}{4(3+2\epsilon)S_{\rm h_1}}\delta^2\bm{q}^2N_c^2{
[1,0,0,1,1,0,0,1,1]};\quad S_{\rm h_1}=1.\\
\left[i{\cal M}_{\rm i_1}\right]_{\rm enh}&=\frac{ig^4}{2S_{\rm i_1}}(\bm{q}^2)^2N_c^2\bigg[\delta^2\bigg\{2\bm{q}^2{
[1,1,1,1,1,1,0,0,1]}+{
[1,1,1,1,0,1,0,0,1]}\\&-4{
[1,1,1,0,1,1,0,0,1]}\bigg\}+8\xi{
[1,1,1,1,1,1,-2,0,1]}\bigg];\quad S_{\rm i_1}=2.\\
\left[i{\cal M}_{\rm j_1}\right]_{\rm enh}&=-\frac{3ig^4}{2S_{\rm j_1}}\bm{q}^2 N_c^2\delta^2\frac{19+12\epsilon}{3+2\epsilon}{
[1,0,0,1,1,0,0,1,1]};\quad S_{\rm j_1}=2.\\
\left[i{\cal M}_{\rm k_1}\right]_{\rm enh}&=\frac{3ig^4}{2S_{\rm k_1}}(\bm{q}^2)^2N_c^2\delta^3{
[1,0,0,1,1,1,1,1,1]};\quad S_{\rm k_1}=6.\\
\left[i{\cal M}_{\rm m_1}\right]_{\rm enh}&=\frac{ig^4}{S_{\rm m_1}}(\bm{q}^2)^2N_c^2\delta\bigg[ -\frac{6\delta}{\bm{q}^2}{
[1,0,0,1,1,0,0,1,1]}+2{
[1,1,1,0,1,1,0,0,1]}\\&+2\xi{
[1,1,0,1,1,1,1,-1,1]}\bigg];\quad S_{\rm m_1}=1.\\
\left[i{\cal M}_{\rm n_1}\right]_{\rm enh}&=0.
\end{aligned}
\end{equation}
In some cases, we have used the Mathematica package FIRE \cite{Smirnov:2008iw} that implements
the Laporta algorithm \cite{Laporta:2001dd}  to reduce the number and
complexity of master integrals through integration-by-parts identities
\cite{Chetyrkin:1981qh}.  Discarding all contributions which are finite or suppressed in the limit   $\rho \to \infty$, we can express the entire  unsubtracted two-loop
self-energy in terms of  7 master integrals $\mathcal{A} - \mathcal{G}$ with a  certain coefficient associated with each master integral, see Tab.~\ref{tab:masters}.
\renewcommand{\arraystretch}{2}
\begin{table}[th]
  \centering
  \begin{align*}
  \begin{array}[h]{rcl|c}
&&\text{master integral} & \text{coefficent} \\
\hline 
   {\cal A}   &\equiv &  \big[1,1,1,1,0,1,0,0,1\big]  & \displaystyle  c_{\mathcal{A}} = -\frac{\bm{q}^2}{2}\\ 
 {\cal B}   &\equiv & 
 \big[1,0,0,1,1,0,0,1,1\big] &  \displaystyle  c_{\mathcal{B}} = \frac{66+42\epsilon}{3+2\epsilon}\\ 
 {\cal C}   &\equiv & 
\big[1,1,1,1,1,1,0,0,1\big]  &  \displaystyle  c_{\mathcal{C}} = -(\bm{q}^2)^2\\ \displaystyle
 {\cal D}   &\equiv & 
[1,0,0,1,1,1,1,1,1\big]  & c_{\mathcal{D}} = -\bm{q}^2 \\
  {\cal E}   &\equiv &
\big[1,1,0,1,1,1,1,-1,1\big] & c_{\mathcal{E}} = -2{
\xi}\bm{q}^2\\
 {\cal F}   &\equiv & 
\big[1,1,1,1,1,1,-2,0,1\big] &     \displaystyle c_{\mathcal{F} = }-{
\xi}\bm{q}^2\\
 {\cal G}   &\equiv &  
\big[1,1,1,0,1,1,0,0,1\big]  &  \displaystyle c_{\mathcal{G}} = 0 \\
  \end{array}
\end{align*}
  \caption{Coefficients of the master integrals. Each coefficient is in addition to be multiplied with the common overall factor  $(-2i\bm{q}^2)g^4N_c^2$.}
  \label{tab:masters}
\end{table}
The master integral ${\cal A}$ can be shown to vanish by symmetry due to
the symmetric pole prescription of the eikonal poles of
the induced vertices. The $\rho$-enhanced pieces of the remaining
master integrals are computed up to terms of order ${\cal O}(\epsilon)$ using the Mellin-Barnes representations technique, for a review see
e.g. \cite{Smirnov:2006ry}.\\

To this end, we first derive multi-contour integral representations
for the master integrals, referring the reader for details to Appendix
\ref{app1}. Having as working environment the code \texttt{MB.m}~\cite{Czakon:2005rk}\footnote{The 
package \texttt{MBresolve.m}~\cite{Smirnov:2009up} was also used.},
we use the Mathematica package \texttt{MBasymptotics.m}~\cite{Cza}
to perform an asymptotic expansion in $e^{-\rho}$. We remove
any terms proportional to $e^{-k\rho\epsilon},\,k\in\mathbbm{Z}$,
capturing this way the leading behavior in
$\rho$. 
As a final step, we resolve the singularities structure
in $\epsilon$ by using the Mathematica packages \texttt{MB.m} and \texttt{MBresolve.m}.
Eventually, some of the final integrals are further simplified by using 
the Barnes' lemmas implemented in the Mathematica code \texttt{barnesroutines.m}~\cite{Kos}.\\

Following this procedure we obtain for  the master integrals the   following results:\footnote{For details on the computation of imaginary parts, see Appendix \ref{app2}.}
\begin{align}
\label{eq:masters_result}
c_{\mathcal{B}} \cdot {\cal B}
&=
\frac{1}{(4\pi)^4}\left[ \frac{11}{\epsilon^2}-\frac{1+66\Xi}{3\epsilon}+\frac{400+12\Xi+396\Xi^2-33\pi^2}{18}\right]\rho, \notag 
\\
c_{\mathcal{C}} \cdot {\cal C}
&=
\frac{1}{(4\pi)^4}\bigg(\bigg[-\frac{4}{\epsilon^3}-\frac{8(1-\Xi)}{\epsilon^2}-\frac{\pi^2+8(1-\Xi)^2}{\epsilon}-   2\pi^2(1-\Xi)  - 
 \frac{16(1-\Xi)^3}{3}\notag  \\
&   -   \frac{50}{3}\zeta(3)\bigg]\rho
+
\left[\frac{2}{\epsilon^2}+\frac{4(1-\Xi)}{\epsilon}+\frac{1}{3}(12(1-\Xi)^2-\pi^2)\right]\left\{\rho^2-i\pi\rho\right\}\bigg), \notag
\\
c_{\mathcal{D}}\cdot{\cal D}
&=
\frac{1}{(4\pi)^4}\bigg[
\frac{4}{\epsilon^3}  
+
\frac{8(1-\Xi)}{\epsilon^2}
+
\frac{4(\pi^2+6(1-\Xi)^2)}{3\epsilon}
+
\frac{8 \pi^2(1-\Xi)}{3}  
\notag \\
&
 +\frac{16 (1-\Xi)^3}{3} +   \frac{44 \zeta(3)}{3}\bigg], \notag
\\
c_{\mathcal{E}} \cdot {\cal E}& =  0,
\notag \\
c_{\mathcal{F}}\cdot {\cal F}& = 
\frac{1}{(4\pi)^4}\left[-\frac{4}{\epsilon^2}+\frac{8\Xi}{\epsilon}+
\frac{2\pi^2}{3} -8 (1+\Xi^2)\right],
\end{align}
where we introduced the notation
\begin{align}
  \label{eq:Xi}
  \Xi&=  1-\gamma_E-\ln \frac{\bm{q}^2}{4\pi\mu^2}.
\end{align}
Using these results, the (unsubtracted)  contribution to the reggeized gluon self-energy (with $n_f = 0$) 
reads:
\renewcommand{\arraystretch}{3}
\begin{align}\label{uns}
\vcenter{\hbox{\includegraphics[scale=0.5]{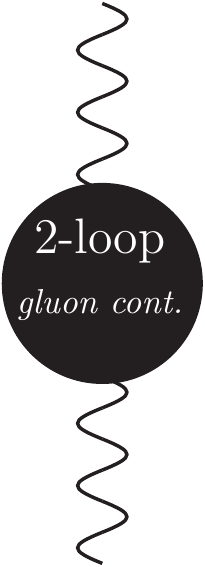}}}
&=
\begin{array}[h]{l}
  \displaystyle 
(-2i{\bm q}^2) \frac{g^4 N_c^2}{(4\pi)^4}\Bigg(\left\{\frac{2}{\epsilon^2}+\frac{4(1-\Xi)}{\epsilon}+
4 (1-\Xi)^2-\frac{\pi^2}{3}\right\}\rho^2 +
\bigg\{ \frac{7}{\epsilon^2} -\frac{14 \Xi}{\epsilon}
\\ \vspace{.cm}
 \displaystyle  \qquad \qquad 
-
 \frac{1-\pi^2}{3\epsilon}  -2 \frac{\Xi(\pi^2-1)}{3}+ 14(1+\Xi^2) + \frac{2}{9} -\frac{\pi^2}{2}-2\zeta(3)  \\
\displaystyle  \qquad \qquad  \qquad 
- i \pi  \left[ \frac{2}{\epsilon^2} + 4\frac{ 1-\Xi}{\epsilon} + \frac{1}{3}(12(1-\Xi)^2-\pi^2)\right]\bigg\}\rho \Bigg)\, . 
\end{array}
\end{align}

 Expanding in $\epsilon$ the expression in Eq.~\eqref{eq:coeff_2loop}, one eventually finds for  the  subtracted reggeized gluon self-energy  for $n_f=0$:
\begin{align}
\label{meister}
\Sigma^{(2)}_{{n_f= 0}}\left(\rho,\frac{\bm{q}^2}{\mu^2}\right)
& =
 \parbox{1cm}{ \includegraphics[height = 2cm]{self_2loop.pdf}}  
  = 
  \parbox{1cm}{ \includegraphics[height = 2cm]{self_2loop_ME.pdf}}
  -
  \parbox{1cm}{ \includegraphics[height = 2cm]{self2_1loop.pdf}}
= (- 2i{\bm q}^2)\frac{g^4 N_c^2}{(4\pi)^4}
\bigg\{
-\bigg[
\frac{2}{\epsilon^2}+\frac{4(1-\Xi)}{\epsilon}
\notag \\
&
+ 4(1-\Xi)^2- \frac{\pi^2}{3}
\bigg]\rho^2
+
\bigg[
 \frac{1}{3\epsilon^2}+\frac{1}{9\epsilon}  +\frac{\pi^2}{3 \epsilon}-\frac{2 \Xi}{3 \epsilon}+\frac{\pi^2(11-12\Xi)}{18}
 \notag \\
 &
 +\frac{16}{27} - \frac{2 }{9} \Xi + \frac{2}{3} \Xi^2 -2\zeta(3)\bigg)\bigg]\rho \bigg \} + \mathcal{O}(\epsilon) + \mathcal{O}(\rho^0).&
\end{align}
Now we can compare our result for the 2-loop self-energy with
the definition of the 2-loop gluon Regge trajectory,
Eq.~\eqref{eq:omega2_defined}. At first we note that all divergent
terms $\sim \rho$ cancel against each other since the terms quadratic in $\rho$ in Eq.~\eqref{meister} cancel precisely the term $[ \rho \omega^{(1)}]^2/2$  in Eq.~\eqref{eq:omega2_defined},  {\it i.e.}
\begin{equation}\label{tr}
\begin{aligned}
(\omega^{(1)})^2\frac{\rho^2}{2}+\frac{\Sigma_{\rho^2}^{(2)}}{(-2i\bm{q}^2)} &=0,
\end{aligned}
\end{equation}
if the first term is expanded up to $\mathcal{O}(\epsilon)$.  Taking the function $f^{(1)}$ in the limit $n_f =0$, the remaining terms  then yield  the 2-loop Regge gluon trajectory for zero flavors,
\begin{equation}
\omega^{(2)}(\bm{q}^2)|_{n_f=0}=\frac{(\omega^{(1)}(\bm{q}^2))^2}{4}\left[\frac{11}{3}+\left(\frac{\pi^2}{3}-\frac{67}{9}\right)\epsilon+\left(\frac{404}{27}-2\zeta(3)\right)\epsilon^2\right],
\end{equation}
which is in complete agreement with the results in the literature
\cite{Fadin:1996tb}. The terms proportional to $n_f$ have been calculated in \cite{Chachamis:2012gh}. With the the  flavor-dependent $\rho$-enhanced terms, the subtracted 2-loop self-energy is given by
\begin{align}
 \Sigma^{(2)}_{n_f} \left(\rho; \epsilon, \frac{{\bm q}^2}{\mu^2}    \right) &=  \frac{\rho( -2i{\bm q}^2) \bar{g}^4 4 n_f}{\epsilon N_c} \frac{ \Gamma^2(2 + \epsilon) }{ \Gamma(4 + 2 \epsilon)}
 \left( \frac{{\bm q}^2}{\mu^2} \right)^{2 \epsilon} 
 \bigg(
\frac{\Gamma^2(1 + \epsilon)}{ \Gamma(1 + 2\epsilon)}\frac{4}{\epsilon}  
\notag \\
& \quad \qquad \qquad \quad \qquad \qquad \qquad \qquad  
-\frac{3 \Gamma(1 - 2\epsilon) \Gamma(1 + \epsilon) \Gamma(1 + 2\epsilon)}{\Gamma^2(1 - \epsilon) \Gamma(1 + 3 \epsilon) \epsilon}\bigg),
   \label{eq:traje_quark_coeff}
\end{align}
and one obtains for the 2-loop Regge gluon trajectory with $n_f$ flavors
\begin{equation}
\omega^{(2)}(\bm{q}^2)=\frac{(\omega^{(1)}(\bm{q}^2))^2}{4}
\left[
\frac{11}{3} - \frac{2 n_f}{3 N_c} +\left(\frac{\pi^2}{3}-\frac{67}{9}\right)\epsilon+\left(\frac{404}{27}-2\zeta(3)\right)\epsilon^2
\right].
\end{equation}

\section{Conclusions and Outlook}\label{5}

In this paper we have presented a derivation of the two-loop gluon
Regge trajectory using Lipatov's effective action and a recently
developed computational scheme, which includes a regularization,
subtraction and renormalization procedure.  Our result is in precise
agreement with earlier results present in the literature and thus
provides a highly non-trivial check of the effective action and our 
proposed computational framework. \\

  From a technical point of view, the
main result of the paper is the computation of the 2-loop
reggeized gluon self-energy. Regularizing high energy divergences by
slightly moving the light-like vectors of the effective action away
from the light-cone, we first demonstrated the suppression of a large
class of diagrams through a scaling argument. The remaining diagrams
were then expressed in terms of seven master integrals, which have
been evaluated using multiple Mellin-Barnes representations.  
Our scheme  introduces a consistent general strategy to deal with
more complex computations, with the hope to easy the path to perform further calculations with  
Lipatov's  high-energy effective action.

\section*{Acknowledgements}

We thank J. Bartels, V. Fadin and L. Lipatov for constant support for many years. 
We acknowledge partial support  by the 
Research Executive Agency (REA) of the European Union under the 
Grant Agreement number PITN-GA-2010-264564 (LHCPhenoNet), 
the Comunidad de Madrid through Proyecto HEPHACOS ESP-1473, 
by MICINN (FPA2010-17747),
by the Spanish Government and EU ERDF funds 
(grants FPA2007-60323, FPA2011-23778 and CSD2007- 00042 Consolider Project CPAN) 
and by GV (PROMETEUII/2013/007).
G.C. acknowledges support from Marie Curie Actions (PIEF-GA-2011-298582). M.H.
acknowledges support from the U.S. Department of Energy under contract
number DE-AC02-98CH10886 and a ``BNL Laboratory Directed Research and
Development'' grant (LDRD 12-034).

\appendix

\section{Appendix}
\label{sec:appendix}

In this appendix we present some  details of  the derivation of Mellin-Barnes representations for the general two-loop master integral considered in this work with propagators to arbitrary powers. The principal tool in this analysis is the formula 
\begin{equation}\label{MB}
\begin{aligned}
\frac{1}{(X_1+\cdots+X_n)^\lambda}&=\frac{1}{\Gamma(\lambda)}\frac{1}{(2\pi i)^{n-1}}\int\cdots\int_{-i\infty}^{+i\infty} dz_2\cdots dz_n\prod_{i=2}^n X_i^{z_i}X_1^{-\lambda-z_2-\cdots-z_n}\\&\times\Gamma(\lambda+z_2+\cdots +z_n)\prod_{i=2}^n \Gamma (-z_i), 
\end{aligned}
\end{equation}
where the contours of integration are such that poles with a $\Gamma(\cdots+z_i)$ dependence are to the left of the $z_i$ contour and poles with a $\Gamma(\cdots-z_i)$ dependencies lie to the right of the $z_i$ contour.\\

\subsection{Mellin-Barnes Representation for Master Integrals without  phases }\label{app1}

We consider the integral
\begin{align}
\label{eq:app1}
\mathcal{S}_1  =\int\frac{d^dk}{(2\pi)^d} & \frac{1}{(-k^2-i0)^C[-(k-q)^2-i0]^D[-(k-l)^2-i0]^E}
\notag \\
&   \times  \frac{1}{(-n_a\cdot k-i0)^{\mu_1}(-n_b\cdot k-i0)^{\mu_2}}, 
\end{align}
where the relation  $n_a\cdot q=n_b\cdot q=0$ is implied. Unlike the general  master integral defined in Eq.~(\ref{eq:MASTER}), the contour of integration is in the following  always defined to lie above the singularities introduced by the light cone denominators. The treatment of alternating descriptions, contained in the functions $g_1$ and $g_2$ is summarized in Appendix \ref{app2}.

Using Schwinger parameters, we can write 
\begin{equation}
\begin{aligned}
\mathcal{S}_1&=\frac{i^{C+D+E+\mu_1+\mu_2}}{\Gamma(C)\Gamma(D)\Gamma(E)\Gamma(\mu_1)\Gamma(\mu_2)}\int_0^\infty\cdots\int_0^\infty d\alpha d\beta d\gamma d\tilde{\delta}d\tilde{\sigma}\,
\\
&
\hspace{4cm} \alpha^{C-1}\beta^{D-1}\gamma^{E-1}\tilde{\delta}^{\mu_1-1}\tilde{\sigma}^{\mu_2-1} \int\frac{d^dk}{(2\pi)^d}e^{i{\cal D}},\\
&  {\cal D}=
\alpha k^2+\beta(k-q)^2+\gamma(k-l)^2+\tilde{\delta} n_a\cdot k+\tilde{\sigma} n_b\cdot k\\
&
\qquad \qquad 
=
(\alpha+\beta+\gamma)k^2+\beta q^2+\gamma l^2-2k\cdot\left(\beta q+\gamma l-\left[\tilde{\delta}\frac{n_a}{2}+\tilde{\sigma}\frac{n_b}{2}\right]\right).
\end{aligned}
\end{equation}
\normalsize With a shift in the momentum integral and introducing
parameters
$\lambda=\alpha+\beta+\gamma,\,\xi=\frac{\beta}{\alpha+\beta},\,\eta=\frac{\gamma}{\alpha+\beta+\gamma};\,\,\tilde{\delta}=2\lambda\delta,\,\tilde{\sigma}=2\lambda\sigma,$
and $x=2(\delta+\sigma),\,y=\frac{\delta}{\delta+\sigma}$, we arrive
at
\begin{align}
  \label{eq:S1processed}
  \mathcal{S}_1= &\frac{i^{C+D+E+\mu_1+\mu_2}}{\Gamma(C)\Gamma(D)\Gamma(E)\Gamma(\mu_1)\Gamma(\mu_2)}\int_0^\infty d\lambda \,\lambda^{C+D+E+\mu_1+\mu_2-1} 
\int_0^\infty dx\, x^{\mu_1+\mu_2-1}
 \notag \\
&
\int_0^1 d\xi\, \xi^{D-1}(1-\xi)^{C-1} 
\int_0^1 d\eta \,  \eta^{E-1}(1-\eta)^{C+D-1}\int_0^1 dy y^{\mu_1-1}(1-y)^{\mu_2-1}
\int\frac{d^dk}{(2\pi)^d}
\notag \\
&
\exp\left[ i\lambda\big( k^2-(1-\eta)^2\xi (1-\xi)\bm{q}^2 -\eta(1-\eta)(1-\xi)(-l^2)  -\eta(1-\eta)\xi[-(l-q)^2]\right.
 \notag  \\
& \left. 
-\eta x[y(-n_a\cdot l)+(1-y)(-n_b\cdot l)-x^2(\Psi y(1-y)+e^{-\rho})\big)\right],
\end{align}
where $\Psi\equiv(1-e^{-\rho})^2$. Performing the integration over momentum and the parameter $\lambda$ we obtain with 
Eq.~(\ref{MB})
\begin{equation}
\begin{aligned}
\mathcal{S}_1&=
\frac{i}{(4\pi)^{d/2}\Gamma(C)\Gamma(D)\Gamma(E)\Gamma(\mu_1)\Gamma(\mu_2)}
\int_0^1 d\xi \,\xi^{D-1}(1-\xi)^{C-1}
\int_0^1 d\eta \, \eta^{E-1} (1-\eta)^{C+D-1}
\\& 
\int_0^\infty dx \, x^{\mu_1+\mu_2- 1}
\int_0^1 dy \,
y^{\mu_1-1}(1-y)^{\mu_2-1}  \, 
\int\cdots\int_{-i\infty}^{+i\infty}\frac{dz_2}{2\pi i}\cdots\frac{dz_7}{2\pi i}\Gamma(-z_2)\cdots\Gamma(-z_7)\\
&
\Gamma(z_2+z_3+z_4+z_5+z_6+z_7+C+D+E+\mu_1+\mu_2-d/2) [(1-\eta)^2\xi(1-\xi)\bm{q}^2]^{z_2}
\\
&
\frac{ [\eta(1-\eta)(1-\xi)(-l^2)]^{z_3} [\eta(1-\eta)\xi(-(l-q)^2)]^{z_4}[\eta xy(-a\cdot l)]^{z_5}[\eta x(1-y)(-b\cdot l)]^{z_6}}{[x^2y(1-y)]^{z_2+z_3+z_4+z_5+z_6+z_7+C+D+E+\mu_1+\mu_2-d/2}[x^2(e^{-\rho})]^{-z_7}},
\end{aligned}
\end{equation}
which allows to perform the integrations over the parameters $\xi, \eta, x $ and $y$.  In some cases integrals of the form
\begin{equation}\label{delta}
\int_0^1 dy\, y^{\alpha-1}(1-y)^{-\alpha-1}=\int_0^\infty dt\,t^{-\alpha-1}=2\pi i\delta\left(\alpha\right)
\end{equation}
appear which allow for the reduction of contour integrals. Eventually, we arrive at
\begin{equation}\label{S1}
\begin{aligned}
\mathcal{S}_1&=\frac{i}{(4\pi)^{d/2}}
\int\cdots\int\frac{dz_2}{2\pi i}\frac{dz_3}{2\pi i}\frac{dz_4}{2\pi i}\frac{dz_5}{2\pi i}\frac{dz_7}{2\pi i}
\frac{\Gamma(-z_2)\Gamma(-z_3)\Gamma(-z_4)\Gamma(-z_5)\Gamma(-z_7)}{\Gamma(-2z_7)} 
\\
&
\frac{\Gamma(2z_{234} +z_5+ 2 C+2D+2E+\mu_1+\mu_2-d)}{\Gamma(C)\Gamma(D)\Gamma(E)\Gamma(\mu_1)\Gamma(\mu_2)}
\Gamma\left(-z_{2347}-C-D-E+\frac{d}{2}\right)
\\
&
\Gamma\left(z_{2345}-z_7+C+D+E+\mu_1-\frac{d}{2}\right)
\Gamma\left(-z_{23457}-C-D-E-\mu_1+\frac{d}{2}\right)
\\
&
\frac{\Gamma(-2z_2- z_{34}-2C-2D-E-\mu_1-\mu_2+d) \Gamma(z_{23} + C) 
\Gamma(z_{24}+D)}{\Gamma(-C-D-E-\mu_1-\mu_2+d)}
\\
&
\Psi^{z_{234}-z_7+C+D+E-d/2}   \left(\bm{q}^2\right)^{z_2}
\left(-l^2\right)^{z_3}
\left(-(l-q)^2\right)^{z_4}\left(-n_a\cdot l\right)^{z_5}\\ &
\left(-n_b\cdot l\right)^{-2z_{234}-z_5-2C-2D-2E-\mu_1-\mu_2+d}(e^{-\rho})^{z_7},
\end{aligned}
\end{equation}
where $z_{ijk\ldots} = z_i + z_j + z_k + \ldots$

In an analogous  way, we can derive the following Mellin Barnes representation,
\begin{equation}\label{S2}
\begin{aligned}
{\mathcal{S}_2}\, & {=\int\frac{d^dk}{(2\pi)^d}\frac{1}{(-k^2-i0)^A[-(k-q)^2-i0]^B(- n_a\cdot k-i0)^{\lambda_1}(-n_b\cdot k-i0)^{\lambda_2}}}
\\
&
=\frac{i\Gamma\left(A+B+\frac{\lambda_1+\lambda_2}{2}-\frac{d}{2}\right)}{2(4\pi)^{d/2}\Gamma(A)\Gamma(B)\Gamma(\lambda_1)\Gamma(\lambda_2)}\frac{\Gamma\left(\frac{d}{2}-A-\frac{\lambda_1+\lambda_2}{2}\right)\Gamma\left(\frac{d}{2}-B-\frac{\lambda_1+\lambda_2}{2}\right)}{\Gamma(d-A-B-\lambda_1-\lambda_2)(\bm{q}^2)^{A+B+\frac{\lambda_1+\lambda_2}{2}-\frac{d}{2}}}
\\
&
\times\int\frac{dz}{2\pi i}\frac{\Gamma(-z)}{\Gamma(-2z)}\Gamma\left(z+\frac{\lambda_1+\lambda_2}{2}\right)\Gamma\left(-z+\frac{\lambda_1-\lambda_2}{2}\right)\Gamma\left(-z-\frac{\lambda_1-\lambda_2}{2}\right)\left(e^{-\rho}\right)^z.
\end{aligned}
\end{equation}
where again $n_a \cdot q = n_b \cdot q = 0$ is implied. 
Iterating the results Eq.~\eqref{S1} and Eq.~\eqref{S2}, we obtain the Mellin Barnes  representation of  the general two-loop master integral
\begin{equation}\label{S}
\begin{aligned}
\mathcal{S}& {=\iint\frac{d^dk}{(2\pi)^d}\frac{d^dl}{(2\pi)^d}\frac{1}{[-k^2-i0]^{\alpha_1}[-(k-q)^2-i0]^{\alpha_2}[-l^2-i0]^{\alpha_3}[-(l-q)^2-i0]^{\alpha_4}}}\\
&  \times {\frac{1}{ [-(k-l)^2-i0]^{\alpha_5} (-n_a\cdot k-i0)^{\alpha_6}(-n_b\cdot k-i0)^{\alpha_7}( n_a\cdot l-i0)^{\alpha_8}(- n_b\cdot l-i0)^{\alpha_9}}}
\\
&=\frac{- \left(\bm{q}^2\right)^{d-{\alpha_{12345}}-\frac{\alpha_{6789}}{2}}}{2(4\pi)^d}
\prod_{i=1}^6 \int\frac{dz_i}{2\pi i} \Gamma(-z_i) \,  \frac{\Gamma\left(z_{1234}+{\alpha_{345}}+\tfrac{\alpha_{6789}}{2}-\frac{d}{2}\right)  \Gamma(-z_4+{\alpha_2}) }{ \prod_{j=1}^9\Gamma(\alpha_j) \Gamma(-2z_1)\Gamma(-2z_6) }
\\
&
\frac{  \Gamma\left(-z_{12345}-{\alpha_{345}}+\tfrac{\alpha_{6789}}{2}+\tfrac{d}{2}\right)\Gamma\left(-z_1+z_{2345}+{\alpha_{345}}-\tfrac{\alpha_{6789}}{2}-\tfrac{d}{2}\right)  \Gamma(-z_{3}+{\alpha_1})  }{\Gamma(-2z_2-z_{34}-{\alpha_{126789}}-2{\alpha_{345}}+2d)  }
\\
&
 \frac{\Gamma\left(-z_{23}-{\alpha_{2345}}-\tfrac{\alpha_{6789}}{2}+d\right)\Gamma\left(-z_{24}-{\alpha_{1345}}-\tfrac{\alpha_{6789}}{2}+d\right)\Gamma(z_{2345}-z_6+{\alpha_{3458}}-\tfrac{d}{2})}{\Gamma(2z_{234}+z_5+2{\alpha_{345}}+\alpha_{789}-d)}
\\
&
\frac{\Gamma(2z_{234}+z_5+2{\alpha_{345}}+\alpha_{89}-d)\Gamma(-z_{234}+z_6-{\alpha_{345}}+\tfrac{d}{2}) \Gamma\left(z_2+{\alpha_{12345}}+\tfrac{\alpha_{6789}}{2}-d\right)}{\Gamma(-z_5+\alpha_6)}
\\
& 
\frac{ \Gamma(-z_{23456}\! -\!{\alpha_{3458}}+\tfrac{d}{2})\Gamma(-2z_2-z_{34}-2\alpha_{34}  -{\alpha_{589}} +d) 
 \Gamma(z_{23}+{\alpha_3})   \Gamma(z_{24}+{\alpha_4})   }
{ 
 \Gamma(-{\alpha_{34589}}  +d)}     e^{-z_{16}\rho}    
\end{aligned}
\end{equation}
where $z_{ijk\ldots} = z_i + z_j + z_k + \ldots$ and $\alpha_{ijk\ldots} = \alpha_i + \alpha_j + \alpha_k + \ldots$. At this stage one then turns to explicit values for the parameters $\alpha_i$, $i = 1, \ldots, 9$ and the integrals are expanded for the limits $\rho \to \infty$ and $\epsilon \to 0$ as explained in Sec.~\ref{sec:calcench}.

\subsection{Computation of $\rho$-enhanced imaginary parts}
\label{app2}

Among all integrals, only the masters $\mathcal{C}$ and $\mathcal{D}$
are, for their $\rho$-enhanced terms, sensitive to the details of the pole prescription. For diagram $(k_1)$, which is directly proportional to
$\mathcal{D}$ and constitutes  the only diagram containing this
master,  explicit QCD calculations  allow to argue that no enhanced imaginary parts can result from such a diagram. This is immediately clear if one identifies this diagram with the high energy expansion of the quark-quark scattering
amplitude with three gluon exchange (see for instance
\cite{Forshaw:1997dc}), which allows to 
 argue that the $\rho$-enhanced imaginary part of this diagram needs to vanish. We verified
that this is indeed the case and we were able to confirm that the entire
$\rho$-enhanced contribution of this master integral coincides with
the equivalent integral using the pole prescription of
Sec.~\ref{app1}.

The master  $\mathcal{C}$ possesses on the other hand a $\rho$-enhanced imaginary part. To this end we consider the integral
 \begin{align}
  \label{eq:C}
  \mathcal{C}^{(\pm, \pm)} & = (\mu^4)^{-2\epsilon} \int \int \frac{d^d k}{(2\pi)^d} \frac{d^d l}{(2\pi)^d}
\frac{1}{[-k^2 - i0][-(k-q)^2 - i0][-l^2 -i0]} \notag \\
& \qquad \qquad \frac{1}{[-(l-q)^2 -i0][-(k-l)^2-i0]} \cdot 
\frac{1}{-n_a \cdot k \pm  i0}\frac{1}{-n_b \cdot k \pm  i0},
\end{align}
where the  integral $\mathcal{C}^{(--)}$ is assumed to be known using the techniques of Sec.~\ref{app1} while  $\mathcal{C}^{(+,+)} = \mathcal{C}^{(-,-)}$ holds.  Introducing rescaled vectors $
  a, b = \frac{1}{2}e^{\rho/2}n_{a,b}$  with $
  a^2  = 1 = b^2 $ and $  a\cdot b  =  \cosh \rho
$
we find
\begin{align}
  \label{eq:CtoD}
   \mathcal{C}^{(\pm, \pm)} & =4 e^{-\rho} \cdot \tilde{C}^{(\pm, \pm)},  \notag \\
\tilde{C}^{(\pm, \pm)} &=  (\mu^4)^{-2\epsilon} \int \int \frac{d^d k}{(2\pi)^d} \frac{d^d l}{(2\pi)^d}
\frac{1}{[-k^2 - i0][-(k-q)^2 - i0][-l^2 -i0]} \notag \\
& \qquad \qquad \frac{1}{[-(l-q)^2 -i0][-(k-l)^2-i0]} \cdot 
\frac{1}{-a \cdot k \pm  i0}\frac{1}{-b \cdot k \pm  i0} \notag \\
&= \frac{ e^{\rho}}{4} C^{(\pm, \pm)}.
\end{align}
As $a^2 = 1 = b^2$, the  new integral is an analytic function of $a\cdot b$ and ${\bm q}^2$  only,  $\tilde{C} = \tilde{C} (a\cdot b, {\bm q}^2)$. With 
\begin{align}
  \label{eq:apparent}
  \frac{1}{-a\cdot k + i0} & = -\frac{1}{-( e^{-i\pi} a)\cdot k - i0},
\end{align}
we have 
\begin{align}
  \label{eq:relationTilde}
   \tilde{C}^{\pm, \mp} (a\cdot b, {\bm q}^2) & = -  \tilde{C}^{(+, + )} ( e^{- i\pi} a\cdot b, {\bm q}^2).
\end{align}
Evaluating all integrals in the limit $\rho \to \infty$, the substitution $ a\cdot b \to  e^{- i\pi} a\cdot b$  is equivalent to a substitution $\rho \to \rho - i\pi$, up to exponentially suppressed corrections. We therefore find
\begin{align}
  \label{eq:Cre}
  \mathcal{C}^{(\mp, \pm)}(\rho) & = \mathcal{C}^{(+,+)}(\rho - i\pi)
\end{align}
which exhausts all possible cases present in Eq.~(\ref{eq:C}).

\pagestyle{plain}

\addcontentsline{toc}{chapter}{\numberline{}\sffamily \bfseries References}


\begin{thebibliography}{99}
\providecommand{\url}[1]{{\tt #1}}
\providecommand{\urlprefix}{ }
\providecommand{\selectlanguage}[1]{\relax}
\providecommand{\eprint}[2][]{\url{#2}}




\bibitem{jets}
  A.~Sabio Vera,
  Nucl.\ Phys.\ B {\bf 746} (2006) 1
  [hep-ph/0602250];
  A.~Sabio Vera and F.~Schwennsen,
  Nucl.\ Phys.\ B {\bf 776} (2007) 170
  [hep-ph/0702158];
  Phys.\ Rev.\ D {\bf 77} (2008) 014001
  [arXiv:0708.0549 [hep-ph]];
  C.~Marquet and C.~Royon,
  Phys.\ Rev.\ D {\bf 79} (2009) 034028
  [arXiv:0704.3409 [hep-ph]];
  M.~Deak, F.~Hautmann, H.~Jung and K.~Kutak,
  JHEP {\bf 0909} (2009) 121
  [arXiv:0908.0538 [hep-ph]];
  M.~Deak, F.~Hautmann, H.~Jung and K.~Kutak,
  Eur.\ Phys.\ J.\ C {\bf 72}, 1982 (2012)
  [arXiv:1112.6354 [hep-ph]];
  M.~Angioni, G.~Chachamis, J.~D.~Madrigal, A.~Sabio Vera;
  Phys.\ Rev.\ Lett.\  {\bf 107} (2011) 191601 [arXiv:1106.6172 [hep-th]].
  
\bibitem{upd}
  H.~Jung, S.~Baranov, M.~Deak, A.~Grebenyuk, F.~Hautmann, M.~Hentschinski, A.~Knutsson and M.~Kramer {\it et al.},
  Eur.\ Phys.\ J.\ C {\bf 70} (2010) 1237
  [arXiv:1008.0152 [hep-ph]];
  H.~Jung and F.~Hautmann,
  arXiv:1206.1796 [hep-ph];
  F.~Hautmann, M.~Hentschinski and H.~Jung,
  Nucl.\ Phys.\ B {\bf 865} (2012) 54
  [arXiv:1205.1759 [hep-ph]];
  A.~V.~Lipatov and N.~P.~Zotov,
  Phys.\ Lett.\ B {\bf 704}, 189 (2011)
  [arXiv:1107.0559 [hep-ph]].



\bibitem{hic}
  B.~Schenke, P.~Tribedy and R.~Venugopalan,
  Phys.\ Rev.\ C {\bf 86}, 034908 (2012)
  [arXiv:1206.6805 [hep-ph]];
  K.~Kutak and S.~Sapeta,
  Phys.\ Rev.\ D {\bf 86}, 094043 (2012)
  [arXiv:1205.5035 [hep-ph]];
  J.~L.~Albacete, A.~Dumitru, H.~Fujii and Y.~Nara,
  Nucl.\ Phys.\ A {\bf 897}, 1 (2013)
  [arXiv:1209.2001 [hep-ph]].



\bibitem{BFKL1}  
L.~N.~Lipatov, 
Sov.\ J.\ Nucl.\ Phys.\  {\bf 23} (1976) 338;
%
E.~A.~Kuraev, L.~N.~Lipatov, V.~S.~Fadin,
Phys.\ Lett.\  B {\bf 60} (1975) 50, 
Sov.\ Phys.\ JETP {\bf 44} (1976) 443,
Sov.\ Phys.\ JETP {\bf 45} (1977) 199;
%
Ia.~Ia.~Balitsky, L.~N.~Lipatov, 
Sov.\ J.\ Nucl.\ Phys.\  {\bf 28} (1978) 822. 

\bibitem{BFKLNLO}
V.~S.~Fadin, L.~N.~Lipatov, Phys.\ Lett.\  B {\bf 429} (1998) 127 [hep-ph/9802290];
%
 M.~Ciafaloni, G.~Camici, Phys.\ Lett.\  B {\bf 430} (1998) 349 [hep-ph/9803389].




\bibitem{Ellis:2008yp}
  J.~Ellis, H.~Kowalski and D.~A.~Ross,
  Phys.\ Lett.\ B {\bf 668} (2008) 51
  [arXiv:0803.0258 [hep-ph]],
  H.~Kowalski, L.~N.~Lipatov, D.~A.~Ross and G.~Watt,
  Eur.\ Phys.\ J.\ C {\bf 70} (2010) 983
  [arXiv:1005.0355 [hep-ph]].
\bibitem{Hentschinski:2012kr}
  M.~Hentschinski, A.~Sabio Vera and C.~Salas,
  Phys.\ Rev.\ Lett.\  {\bf 110} (2013) 041601
  [arXiv:1209.1353 [hep-ph]],
Phys.\ Rev. D\ {\bf 87} (2013) 076005
  [arXiv:1301.5283 [hep-ph]].


\bibitem{Dusling:2012cg} 
  K.~Dusling and R.~Venugopalan,
  Phys,\ Rev.\ D\ {\bf 87} (2013) 051502 [arXiv:1210.3890 [hep-ph]].

\bibitem{Colferai:2010wu}
  D.~Colferai, F.~Schwennsen, L.~Szymanowski and S.~Wallon,
  JHEP {\bf 1012} (2010) 026
  [arXiv:1002.1365 [hep-ph]],
\bibitem{Ducloue:2013hia} 
  B.~Duclou\'e, L.~Szymanowski and S.~Wallon,
  JHEP {\bf 1305}, 096 (2013)
  [arXiv:1302.7012 [hep-ph]].


\bibitem{Caporale:2012ih}
  F.~Caporale, D.~Yu.~Ivanov, B.~Murdaca and A.~Papa 
 [arXiv:1211.7225 [hep-ph]],
  F.~Caporale, B.~Murdaca, A.~Sabio Vera and C.~Salas,
  arXiv:1305.4620 [hep-ph].


\bibitem{LevSeff}
  L.~N.~Lipatov,
  Nucl.\ Phys.\  {\bf B452 } (1995)  369 [hep-ph/9502308],
  Phys.\ Rept.\  {\bf 286 } (1997)  131 [hep-ph/9610276].
\bibitem{Antonov:2004hh} 
  E.~N.~Antonov, L.~N.~Lipatov, E.~A.~Kuraev and I.~O.~Cherednikov,
  Nucl.\ Phys.\ B {\bf 721}, 111 (2005)
  [hep-ph/0411185].




\bibitem{quarkjet} 
  M.~Hentschinski and A.~Sabio~Vera,
  Phys.\ Rev.\ D {\bf 85}, 056006 (2012)
  [arXiv:1110.6741 [hep-ph]].
\bibitem{gluonjet} 
  G.~Chachamis, M.~Hentschinski, J.~D.~Madrigal and A.~Sabio~Vera,
  Phys.\ Rev.\ D\ {\bf 87}, 076009 [arXiv:1212.4992 [hep-ph]].


\bibitem{Fadin:1996tb}
  V.~S.~Fadin, R.~Fiore, M.~I.~Kotsky,
  Phys.\ Lett.\  {\bf B387} (1996)  593 [hep-ph/9605357].

\bibitem{Fadin:1995km} 
  V.~S.~Fadin, R.~Fiore and A.~Quartarolo,
  Phys.\ Rev.\ D\ {\bf 53}, 2729 (1996)
  [hep-ph/9506432];
  M.~I.~Kotsky and V.~S.~Fadin,
  Phys.\ Atom.\ Nucl.\  {\bf 59}, 1035 (1996)
  [Yad.\ Fiz.\  {\bf 59N6}, 1080 (1996)];
  V.~S.~Fadin, M.~I.~Kotsky and R.~Fiore,
  Phys.\ Lett.\ B {\bf 359}, 181 (1995).




\bibitem{Blumlein:1998ib} 
  J.~Bl\"umlein, V.~Ravindran and W.~L.~van Neerven,
  Phys.\ Rev.\ D {\bf 58}, 091502 (1998)
  [hep-ph/9806357].

\bibitem{Korchemskaya:1996je} 
  I.~A.~Korchemskaya and G.~P.~Korchemsky,
  Phys.\ Lett.\ B {\bf 387}, 346 (1996)
  [hep-ph/9607229].
\bibitem{DelDuca:2001gu} 
  V.~Del Duca and E.~W.~N.~Glover,
  JHEP {\bf 0110}, 035 (2001)
  [hep-ph/0109028].


\bibitem{Chachamis:2012gh} 
  G.~Chachamis, M.~Hentschinski, J.~D.~Madrigal and A.~Sabio Vera,
  Nucl.\ Phys.\ B {\bf 861}, 133 (2012)
  [arXiv:1202.0649 [hep-ph]].




\bibitem{review} 
  G.~Chachamis, M.~Hentschinski, J.~D.~Madrigal and A.~Sabio~Vera
  [arXiv:1211.2050 [hep-ph], to appear in Phys. Part. Nucl.].

\bibitem{Hentschinski:2011xg} 
  M.~Hentschinski,
  Nucl.\ Phys.\ B {\bf 859}, 129 (2012)
  [arXiv:1112.4509 [hep-ph]].

\bibitem{Fadin:1998sh} 
  V.~S.~Fadin,
  ``BFKL news,''
  hep-ph/9807528,
  B.~L.~Ioffe, V.~S.~Fadin and L.~N.~Lipatov,
  ``Quantum chromodynamics: Perturbative and nonperturbative aspects,''
  Cambridge monographs on Particle Physics, Nuclear Physics and Cosmology (No.~30)



\bibitem{Bartels:2008ce} 
  J.~Bartels, L.~N.~Lipatov and A.~Sabio Vera,
  Phys.\ Rev.\ D {\bf 80}, 045002 (2009)
  [arXiv:0802.2065 [hep-th]].



\bibitem{Smirnov:2008iw}
  A.~V.~Smirnov,
  JHEP {\bf 0810} (2008) 107 [arXiv:0807.3243 [hep-ph]]. 


\bibitem{Laporta:2001dd}
  S.~Laporta,
  Int.\ J.\ Mod.\ Phys.\  A {\bf 15} (2000) 5087  [hep-ph/0102033]. 


\bibitem{Chetyrkin:1981qh} 
  K.~G.~Chetyrkin and F.~V.~Tkachov,
  Nucl.\ Phys.\ B {\bf 192}, 159 (1981).


\bibitem{Smirnov:2006ry} 
  V.~A.~Smirnov, \emph{Feynman Integral Calculus}. Springer, Berlin (2006).

\bibitem{Czakon:2005rk} 
  M.~Czakon,
  Comput.\ Phys.\ Commun.\  {\bf 175}, 559 (2006)
  [hep-ph/0511200].

\bibitem{Smirnov:2009up} 
  A.~V.~Smirnov and V.~A.~Smirnov,
  Eur.\ Phys.\ J.\ C {\bf 62}, 445 (2009)
  [arXiv:0901.0386 [hep-ph]].
  
\bibitem{Cza}
{ M.~Czakon}, {\em \texttt{MBasymptotics.m}\/},
  \urlprefix\url{http://projects.hepforge.org/mbtools/.}


\bibitem{Kos}
{ D.~A. Kosower}, {\em \texttt{barnesroutines.m}\/}
  \urlprefix\url{http://projects.hepforge.org/mbtools/}.

\bibitem{Forshaw:1997dc} 
  J.~R.~Forshaw and D.~A.~Ross,
  Cambridge Lect.\ Notes Phys.\  {\bf 9}, 1 (1997).
\end{thebibliography}
\end{document}